\documentclass[fullpaper,preprint]{nldl}

\paperID{40}

\vol{V}

\usepackage{mathtools}
\usepackage{amssymb}

\usepackage{graphicx}
\usepackage{tikz}
\usetikzlibrary{positioning, arrows.meta, calc}

\usepackage{booktabs}

\usepackage{enumitem}

\usepackage{algorithm}
\usepackage{algorithmic}

\usepackage{listings}
\usepackage{hyperref}
\usepackage{url}
\hypersetup{
  pdfusetitle,
  colorlinks,
  linkcolor = BrickRed,
  citecolor = NavyBlue,
  urlcolor  = Magenta!80!black,
}

\title{Risk-Sensitive Reward Composition for Conditional GFlowNets}
\author[1]{Carine Ribeiro dos Santos\thanks{Corresponding Author.}}
\author[2]{Ina Pöhner}
\affil[1]{Instituto de Química, Departamento de Química Orgânica, Universidade Federal do Rio de Janeiro}
\affil[2]{School of Pharmacy, University of Eastern Finland}
\affil[ ]{\texttt{carineribeirost@pos.iq.ufrj.br}}

\begin{document}
\maketitle

\begin{abstract}
Generative Flow Networks (GFlowNets) for structure-based drug design condition
on one rigid protein structure. A flexible target holds several distinct
shapes, its \emph{conformations}, each occupied for a fraction of the
simulation time. Scoring a candidate against all of them leaves an open
question: how do \(K\) scores become one reward? Simulation estimates how
often each conformation is occupied, and those estimates are imprecise. Some
conformations must be bound and others must never be, so a candidate that
fails one of them is disqualified however well it scores elsewhere. No
standard aggregation rule handles the imprecision and the disqualification
together.

We build the reward from two quantities. A conditional value-at-risk at tail
level \(\beta\) averages only the worst \(\beta\)-fraction of the \(K\)
scores, so one strong score cannot offset a failed one. An ambiguity radius
\(\rho\) widens the set of population estimates the reward must survive.
Together they index a family of target distributions, and we train one
conditional GFlowNet on the whole family. The synthetic worlds we test on are
small enough to list, so we can write down each target in full and compute the
sampler's error exactly.

The composed rule is one defensible point on a response surface. Crossing it
raises the probability mass on candidates that satisfy every conformation
tenfold on our hardest geometry, and costs 31\% of the distinct candidates.
Set against a hard minimum, a weighted geometric mean and Derringer-Suich
desirability under matched exclusions, what separates it is mainly the number
of distinct candidates it keeps, since it never lets one score offset another.
One network covers the family to within \(0.39\)--\(2.85\) times a
finite-sample reference, the noise floor at our evaluation budget. The floor
construction leaves the reward exactly flat over most of one geometry, and
there on-policy training settles into a degenerate optimum that any off-policy
source prevents. Among those sources, what matters is where the training
signal lands: a regime meeting 99.7\% of the satisfying candidates during
training recovers fewer of them than one meeting 38.5\%.
\end{abstract}

\section{Introduction}
\label{sec:intro}

Generative Flow Networks (GFlowNets) sample objects with probability
proportional to a reward \citep{bengio2021,bengio2021a}. A trained sampler
therefore returns many distinct acceptable candidates, which suits design
problems where diversity matters as much as the score itself. Objects are assembled piece by 
piece, and training enforces consistency between the policy and the reward
\citep{malkin2022,madan2022,malkin2022a,tiapkin2023}. The reward is always a
single number handed to the sampler, so how that reward is \emph{composed} from
several imperfect signals has so far not been examined.

Molecular GFlowNets for drug discovery score a candidate against one target 
protein structure \citep{cretu2025,koziarski2024,shen2025}. Real targets 
hold several distinct shapes, their conformations, and binding selects among a 
pre-existing ensemble \citep{boehr2009,zhou2010}. For dynamic targets, this is often modelled
computationally by docking to each representative \citep{amaro2008}. 
Some conformations must be bound and others avoided, so binding the dominant
one is often not enough. Conformational populations are estimated
with error, so a fixed weight vector would be misleading. A candidate that fails a required conformation is disqualified, and no standard aggregation rule expresses this: a weighted sum allows one score to compensate for another, a Pareto front retains the disqualified candidate on the frontier, and an action mask must act before the disqualifying scores exist.
Desirability functions \citep{derringer1980} come closest, encoding
disqualification multiplicatively, but they fix the aggregation shape in advance and
offer no account of weight uncertainty.

We compose the reward from two robust optimisation quantities. Conditional 
value-at-risk (CVaR) \citep{rockafellar2000} at tail level \(\beta\) averages only the
worst \(\beta\)-fraction of the scores. An ambiguity ball of radius
\(\rho\) around the stated weights \citep{bental2013}, the standard
Distributionally Robust Optimisation (DRO) ambiguity set, widens the set of
weightings the reward must survive. Together with a temperature \(\beta_t\) and auxiliary weight \(w_g\), these
four \emph{control parameters} define a condition \(c\) and a fixed target
\(p^*_c \propto R(x)^{\beta_t}\) before training. We reserve the term 
\emph{control parameters} for
quantities a practitioner sets on the reward, as opposed to the network's
learned parameters which training fits, and the constants that define a
synthetic world, which we draw. We abstract the requirements into four constraint geometries
(including a floor and veto) modelled on multimeric proteins. Such proteins are composed of multiple subunits
and a molecule may need to bind each subunit, avoid an interface, or clear a
bulk threshold.

The targets induced by these four control parameters cannot be obtained by sweeping the temperature
of a single target, since the risk parameters reorder candidates while
temperature does not (Appendix~\ref{app:reorder}). The argument concerns the
targets alone, and it leaves open whether some other conditioning mechanism
could reach the same family. We ask four questions: does the aggregation change the target
enough to justify conditioning? Can one sampler learn the entire family? What
does it cost? Where does it break? We answer on fully enumerable synthetic
worlds, where target, policy, and exact \(L_1\) distance are computable in
closed form. Extending the standard enumerable-grid testbed
\citep{kim2024,kim2026}, our worlds are large enough for evaluation across a
continuous family of conditions, and an exact-Kullback-Leibler (KL) oracle (a
copy of the network trained on the true target) shows what cross-entropy
training reaches when the target is known. Because the scores here are analytic, the results speak to the construction of the reward and to the training of the sampler.

Our contributions are:

\begin{itemize}[leftmargin=*, itemsep=1pt, topsep=2pt]
  \item \textbf{A reward composition for an ensemble of conformations.} We
    combine a CVaR tail level \(\beta\) with an ambiguity radius \(\rho\) on
    the stated weights, and add four constraint geometries for the
    requirements that disqualify a candidate outright. The combination fixes a
    target \(p^*_c \propto R(x)^{\beta_t}\) before any training
    (\S\ref{sec:targets}).

  \item \textbf{A response surface for the composed reward, and a comparison
    against three other conjunctive rules.} We compute the exact target on an
    \(11 \times 9\) grid of \((\beta, \rho)\) for eight worlds per case, and
    we score a hard minimum, a weighted geometric mean and Derringer-Suich
    desirability on the same worlds under matched exclusions. The surface
    places our rule at one point on it, and the comparison separates it from
    the three alternatives mainly by how many distinct candidates it keeps,
    given that it never lets one score offset another
    (\S\ref{sec:aggregation}).

  \item \textbf{A measurement of what amortising the family costs.} We train
    one conditional network per world and seed, score it by exact held-out
    \(L_1\) on all four cases at deployment scale, and read it against two
    references trained on the same worlds: per-condition experts, which serve
    a single condition each, and an exact-KL oracle fitted by cross-entropy to
    the enumerated target. One network stays within \(0.39\)--\(2.85\) times
    the finite-sample reference, and the experts reach \(1.4\) to \(3.4\)
    times lower error at matched compute (\S\ref{sec:learnability}).

   \item \textbf{A measurement of how held-out error scales along four
    difficulty axes}: space size, depth, weight concentration, and score noise.
    We also measure which of these a larger training budget mitigates, and find
    that tripling the budget recovers the depth axis. Extreme weight imbalance
    does not improve with more training, and conditioning on the score-noise
    margin breaks the endpoint even for the oracle (\S\ref{sec:price}).

  \item \textbf{A controlled comparison of five off-policy training regimes},
    separated by whether training states are injected from outside the
    policy's support or reweighted from states it has already visited.
    Measuring the trained policy, the regimes separate first by whether any
    off-policy source is present at all. Among those that have one, what
    matters is where the training signal lands. The floor construction leaves a large region of exactly constant reward, in which an on-policy learner can settle at a degenerate optimum, and that same region absorbs most of the budget of a uniform off-policy half (\S\ref{sec:sparsity}).

  \item \textbf{An exact testbed.} Because the worlds are small enough to
    list, we compute the target, the policy's terminating distribution and the
    \(L_1\) distance between them in closed form across a continuous family
    of conditions. Every error we report is thus exact (\S\ref{sec:testbed},
    \S\ref{sec:learnability}).
\end{itemize}

\section{Related work}
\label{sec:related}

\paragraph{Aggregation in multi-objective GFlowNets.}
Combining several quality scores into one number predates GFlowNets: the
desirability-function tradition \citep{derringer1980} builds a non-linear
aggregate whose geometric form already encodes that a strong response cannot 
rescue a failed one, and the multi-objective molecular optimisation
literature surveys the trade-offs at length \citep{fromer2022}. Within
GFlowNets, the prevailing choice is linear or rank-based. MOGFN
\citep{jain2022a} and HN-GFN \citep{zhu2023} condition on a single preference
vector or hypernetwork-generated weights, and goal-conditioned MOGFN variants
amortise over desired outcome regions \citep{roy2023}. Order-Preserving
\citep{chen2023} and Global-Order \citep{pastorperez2025} GFlowNets avoid
weights via Pareto-consistent rewards. All place the conditioning
variable at a \emph{point} of the weight simplex (or discard weights entirely),
whereas we place it on a \emph{set} and score a candidate by the worst reweighting inside it. The only
non-linear aggregation in the literature to date is EP-GFlowNet's product form
\citep{silva2024}, which is fixed by its federated setting and cannot be
tuned by a user. The problem is not confined to GFlowNets. RePO \citep{li2026} rewards
molecular optimisation with an unweighted sum of a binary improvement
indicator and a Tanimoto similarity in \([0,1]\), so a molecule that fails
the property outright can still score up to 1.0 on similarity alone and tie
with one that improves the property but shares no similarity with the reference. That is the offset
our composition is built to refuse. None of
this prior work applies a risk measure to the \emph{tail} of the score distribution
under weight uncertainty.

\paragraph{Risk-sensitivity in GFlowNets.}
Risk-sensitivity has appeared in GFlowNets, but applied to stochastic rewards,
not deterministic score aggregation. Quantile flows
\citep{bellemare2017,dabney2018,zhang2023} apply distortion measures (e.g.,
CVaR) to terminal reward noise, inheriting a convention from risk-sensitive 
reinforcement learning
in which the risk level belongs to the run and the user does not supply it
\citep{chow2015,tamar2015}. Our aggregation is deterministic once the scores exist, and we supply the
risk parameters as conditioning, so one network serves a range of
preferences. The closest related work sits in
continuous-flow fine-tuning: Flow Density Control \citep{santi2025} and
Tail-aware Flow Fine-Tuning \citep{wang2026} optimise risk-averse objectives,
but they maximise a risk functional of the distribution the policy
\emph{induces}, which leaves the target defined only as the solution of an
optimisation problem. We apply the risk measure inside the reward instead, so
the target is a fixed distribution that exists before training. The ambiguity
layer, a ball around the stated weights, has no precedent in GFlowNets, though
it is standard in DRO \citep{delage2010,bental2013,rahimian2022,duchi2021}. The
CVaR-DRO correspondence \citep{rockafellar2000,bental2007} and the tilted-loss
view connecting both to exponential reweighting \citep{li2021} are why we can
use a tail level and a radius as two control parameters on a single
construction. Both
act on a coherent risk measure \citep{artzner1999}, which enables the
reordering argument of Appendix~\ref{app:reorder}.

\paragraph{Generation against molecular targets.}
Prior work on target-specific molecular generation assumes a fixed protein
pocket. This includes autoregressive placement in a fixed pocket
\citep{luo2021}, joint ligand-pose diffusion \citep{schneuing2024}, and flow
matching \citep{cremer2025}. Other work applies GFlowNets through action spaces
that enforce synthesisability \citep{cretu2025,koziarski2024,seo2025,kim2026},
but all fix a single pocket for the whole run. More relevant is conditioning on
the pocket \citep{shen2024taco,shen2025}: it selects \emph{which} target to
bind, whereas our implementation specifies \emph{how} to pool scores against a fixed set of conformations. The two are orthogonal, and a sampler could carry both.

\paragraph{Conditioning and exact evaluation.}
Conditioning a sampler on a description of its own target is established.
Temperature-conditional GFlowNets \citep{kim2023a} sweep the reward exponent,
and Logit-GFN fixes a pathology where different temperatures cause different
logit magnitudes. Goal- and outcome-conditioned variants \citep{he2024,pan2023}
amortise over outcomes. Our condition differs in the family it indexes rather
than in the mechanism. For hard constraints, the prevailing approach makes
infeasible objects unreachable by construction \citep{cretu2025,
hernandezgarcia2023,samanta2025}. Our floors and vetoes are evaluated after the candidate exists, and therefore cannot be enforced by making infeasible objects unreachable. Our evaluation also follows
papers that avoid Monte Carlo proxies \citep{shen2023,atanackovic2024,
krichel2024}. \citet{lei2026} provide a formal argument why a loss curve
cannot substitute for a reported distance. We add exact evaluation on
\emph{held-out} conditions and an exact-KL oracle, a separate copy trained
directly on the enumerated target, used solely to attribute residual error.

\paragraph{Exploration under sparsity.}
Exploration in GFlowNets has been addressed through replay and prioritised
variants \citep{vemgal2023,ikram2024}, target networks that supply off-policy
trajectories \citep{lau2023}, local search \citep{kim2023}, uncertainty-driven
novelty \citep{pan2022,rectorbrooks2023}, sequential residual learning
\citep{dallantonia2025}, and adaptive teachers \citep{kim2024,malek2025}. These
are usually compared by sample efficiency or final mode count. None distinguishes
methods by \emph{where the training states come from}: whether injected from
outside the policy's current support or reweighted from states already visited.
That is the axis along which our results split (\S\ref{sec:sparsity}), and
it cuts across the usual taxonomy: local search and teachers are injection,
replay and contrastive replay are reweighting, however elaborate the selection
rule. RewardFlow \citep{feng2026} sits on the reweighting side of the same
axis, propagating topology-aware credit over a state graph built from sampled
trajectories to densify a sparse terminal reward among states the policy has
already reached. Our coverage measurements bear on that choice: with terminal
rewards held fixed and only the source varied, reweighting reaches both the
higher coverage and the lower \(L_1\) on the floor case.

\section{Methods}
\label{sec:methods}

\begin{figure*}[tb]
  \centering
  \begin{tikzpicture}[
    font=\small,
    node distance=8mm and 7mm,
    box/.style={draw=black!70, rounded corners=2pt, align=center,
                inner sep=4pt, minimum height=13mm, text width=23mm},
    arr/.style={-{Stealth[length=2mm]}, semithick, black!70},
    lbl/.style={font=\footnotesize\itshape, text=black!55, inner sep=2pt}
  ]
    \node[box, fill=blue!7] (world) {World $\mathcal{W}$\\[1pt] scores $a$, weights $w$};
    \node[box, fill=orange!10, below=of world] (cond) {Condition $c$\\[1pt] $(\beta_t, w_g, \beta, \rho, \delta_v)$};
    \node[box, fill=gray!8, right=11mm of $(world.east)!0.5!(cond.east)$] (risk)
      {Exact risk layer\\[1pt] $\Psi(x)$: DRO-CVaR\\[1pt] cases A--D};
    \node[box, fill=gray!8, right=of risk] (target)
      {Target\\[1pt] $p^*_c \propto R(x)^{\beta_t}$\\[1pt] all of $\mathcal{X}$};
    \node[box, fill=green!9, right=of target] (gfn)
      {Conditional GFlowNet\\[1pt] FiLM $P_F(x \mid c)$\\[1pt] head $\log Z(c)$};
    \node[box, fill=red!7, right=of gfn] (eval)
      {Exact $L_1$\\[1pt] held-out $c$\\[1pt] $\leq 3\times$ reference};
    \draw[arr] (world.east) to[out=0, in=155] (risk.west);
    \draw[arr] (cond.east)  to[out=0, in=205] (risk.west);
    \draw[arr] (risk) -- (target);
    \draw[arr] (target) -- (gfn);
    \draw[arr] (gfn) -- (eval);
    \draw[arr] (cond.south) -- ++(0,-7mm) -| node[pos=0.24, below, lbl] {condition} (gfn.south);
    \draw[arr] (target.north) -- ++(0,7mm) -| node[pos=0.24, above, lbl] {same enumeration} (eval.north);
  \end{tikzpicture}
  \caption{The exact testbed. A world $\mathcal{W}$ defines score fields and weights; a condition $c$ specifies the risk control parameters. The exact risk layer $\Psi(x)$ maps a candidate's scores to an aggregate, which defines the target $p^*_c$ over all states $\mathcal{X}$. The conditional GFlowNet learns to sample from this target and evaluation computes the exact $L_1$ distance between the learned policy and the target. $\Psi(x)$ is the DRO-CVaR aggregation and the policy is modulated by Feature-wise Linear Modulation (FiLM, \citealp{perez2018}).}
  \label{fig:schematic}
\end{figure*}

\subsection{The exact testbed}
\label{sec:testbed}
All experiments run on synthetic \emph{worlds} over enumerable
\(\mathcal{X} = \{0,\dots,H-1\}^d\), with \(H\) the alphabet size and \(d\) the
length. A \emph{candidate} \(x\) is one point of \(\mathcal{X}\), and the sampler
must produce candidates in proportion to their reward. We use two families: the \emph{calibration} family (\(H=32\), \(d=2\),
\(|\mathcal{X}|=1024\)) fixes the operating point and the \emph{deployment-scale}
sequence family (\(H=4\), \(d=8\), \(|\mathcal{X}|=65{,}536\)) verifies transfer
and studies discovery under sparsity. In the sequence family, each state is a
string of length \(d\) over an alphabet of size \(H\). At \(H=4\) and \(d=8\)
this reproduces the shape of the standard TF-binding sequence benchmark, though
the score fields are synthetic. We draw them as a position-weight matrix plus sparse pairwise epistasis, and with
no binding data behind them. Sequences suffice here because the risk layer
acts on \(a(x) \in [0,1]^K\) alone, so a molecular graph would not alter what it sees.

A testbed need only preserve the scores' statistical structure: bounded,
non-separable in the upper tail, and over a space small enough to enumerate.
The \emph{upper tail} of a score field is its high-scoring end, the candidates
above a high quantile of that field over \(\mathcal{X}\).
\emph{Non-separability} is meant in the standard epistasis sense, that a
candidate's score is not a sum of independent per-position contributions, so
the tail cannot be reached by maximising each position on its own. The
epistatic terms ensure it, since a pure weight matrix is additive and would
make the upper tail a product of per-position maxima. The \(4^8\)
space satisfies all three. A realistic ligand space satisfies the first two and
fails the last, which is a prerequisite for exact evaluation. Scaling variants at
\(H=16, d=3\), at \(H=8, d=4\), and at \(H=64\) covers the size axis.

Enumerability is the design principle (Figure~\ref{fig:schematic}). For any
condition \(c\), both the target \(p^*_c\) and the policy's terminating
distribution \(p_\theta(\cdot \mid c)\) are computable in closed form.
Every error we report is therefore an exact distance, with no Monte-Carlo
estimation anywhere in it.

A world bundles \(K\) bounded \emph{score fields}, with one per conformation. A
score field \(a_k\) is a function that assigns every candidate \(x\) a score
in \([0,1]\) for conformation \(k\). We write
\(a(x) \in [0,1]^K\) for the vector of all \(K\) and \(w\) for their nominal Dirichlet
weights (Dirichlet(2.0); a \emph{peaked} family uses 0.3, where one conformation
dominates). Alongside them sits a
single auxiliary objective \(g\), a score that enters the reward outside the
risk aggregation. On the sequence family, the score fields are generated from a
Position-Weight Matrix (PWM, \citealp{stormo2000}), a standard model for DNA
binding specificity. The sequence-design setting follows GFlowNet work on
biological sequences \citep{jain2022}, combined with a PWM epistasis model for position
interactions, giving a non-separable upper tail for the aggregation to contend
with. On the grid family, the fields are low-frequency mixtures, serving as a
smooth control. The fields are analytic by construction, since substituting a real
scorer would forfeit the exact target our measurement rests on.

Two additional control parameters shape a world. The \emph{sparsity exponent}
\(s \ge 1\) raises each objective field to the power \(s\), which thins its
upper tail, so satisfying candidates become rarer, with \(s = 1\) being the identity
and the constraint thresholds exempt. The \emph{score-robustness margin}
\(\sigma\) is a worst-case allowance for error in the scores themselves: before
aggregation, it lowers every score in the promote set by \(\sigma\) and raises
every score in the suppress set by \(\sigma\).
Worlds are rejection-sampled against a gate: risk-on and risk-off targets must
differ by Total Variation \(\mathrm{TV} \geq 0.05\), the floor of Case B must
fail between 5\% and 70\% of the space, and the vetoes of Case C must cover
between 2\% and 50\% of it. Appendix~\ref{app:gate} reports how often the gate
rejects a world.

Enumeration is needed only to evaluate. The reward is always
computable: a deployment runs \(K\) scorers and evaluates the closed-form dual
of Eq.~\eqref{eq:dro}. Training needs only \(\log R(x)\), whose cost is
independent of \(|\mathcal{X}|\). What a deployment cannot compute is
\(Z_c = \sum_{x} R(x)^{\beta_t}\), which requires enumeration. The synthetic
worlds make \(Z_c\) reachable, so evaluation is a true distance. Since exact
\(Z_c\) is unavailable in deployment, \S\ref{sec:sparsity} also reports an
online diagnostic that needs no enumeration.

\subsection{Risk-sensitive targets}
\label{sec:targets}

Risk enters through a distributionally robust CVaR with two control parameters. The tail
level \(\beta \in (0,1]\) controls how far into the lower tail we look
(\(\beta=1\) recovers the ordinary weighted average). The ambiguity radius
\(\rho\) around the nominal weights expresses how much we distrust them
(\(\rho=0\) trusts them exactly). The aggregation is the DRO-CVaR functional,
defined as
\begin{equation}
  \Phi^-(a; w, \beta, \rho) = \inf_{q \in B_\rho(w)} \mathrm{CVaR}^{-}_{\beta}(a; q),
  \label{eq:dro}
\end{equation}
where the infimum runs over every reweighting \(q\) of the \(K\) signals lying
within the ball \(B_\rho(w)\). The infimum over the ambiguity set \(B_\rho(w)\) is the DRO step, and it makes
the CVaR hold against any reweighting the ball allows. The \(\Phi^+\) variant
is \(\Phi^+(a) = -\Phi^-(-a)\).

The pair \((\beta, \rho)\) interpolates between two extremes. The
\textbf{Boltzmann pole} \((\beta=1, \rho=0)\) is the ordinary weighted average,
which corresponds to the standard temperature-conditional
sampling setup. The \textbf{pure worst case} \((\beta \to 0, \rho \to \infty)\)
scores a candidate by its very worst scores under the most adversarial
reweighting. We
implement three ball geometries \(B_\rho\): KL divergence
(entropic dual), TV (exact greedy), and modified chi-squared
(\(\chi^2\)). For the \(\chi^2\) ball the Karush-Kuhn-Tucker (KKT) conditions
have a closed-form solution on each candidate support. After sorting the
scores, every admissible support is checked, so the worst case is found
exactly. Each ball has its own radius grid, and Appendix~\ref{app:supp} gives
the three duals. All are evaluated outside the learning loop.

The aggregation is deterministic: given scores \(a(x)\) and weights \(w\), the
dual returns a fixed number. No reward noise enters anywhere. The weights themselves are uncertain, and \(\rho\) states how uncertain. Herein, we sweep
\(\rho\) as a design variable, while a deployment would calibrate it from data. Since
the calibrated value differs from target to target, a sampler has to serve a
\emph{range} of radii.

\begin{table*}[tb]
  \centering
  \caption{The four aggregation cases. \(\Phi^{\pm}\) is the DRO-CVaR functional from Eq.~\eqref{eq:dro}. Each score set has its own control parameters \((\beta,\rho)\). The sets are neutral \(S^0\), promote \(S^+\), suppress \(S^-\), and origins \(o = 1..O\) with outer weights \(\pi\). The third column gives the requirement each geometry encodes, together with the class of target it was abstracted from. The experiments use the mathematical form alone.}
  \label{tab:cases}
  \small
  \begin{tabular}{@{}p{1.6cm}p{7.0cm}p{6.6cm}@{}}
    \toprule
    Case & \(\Psi(x)\) & requirement encoded \\
    \midrule
    A smooth & \(\Phi^-(a_{S^0})\) & bind all shapes; none privileged \citep{ferreira2021} \\
    \addlinespace[3pt]
    B floor  & \(\Phi^-(a_{S^+}) - \gamma\, \Phi^+(a_{S^-})\); \newline \(\Phi^-(a_{S^+}) < f \Rightarrow R = \epsilon\) & clear a bar overall, avoid a second set \citep{diazholguin2022} \\
    \addlinespace[3pt]
    C veto   & \(\Phi^-(a_{S^+})\); \newline \(\exists k\!: a_k \geq c_k - \delta_v \Rightarrow R = \epsilon\) & never bind certain shapes at all \citep{kholodenko2015} \\
    \addlinespace[3pt]
    D nested & \(\Phi^-_{\text{out}}\bigl(\Phi^-_{\text{in}}(a_{o,\cdot})_{o=1..O};\, \pi\bigr)\) & pool within groups of shapes, then across \citep{maltarollo2022} \\
    \bottomrule
  \end{tabular}
\end{table*}

\begin{figure*}[tb]
  \centering
  \includegraphics[width=0.86\textwidth]{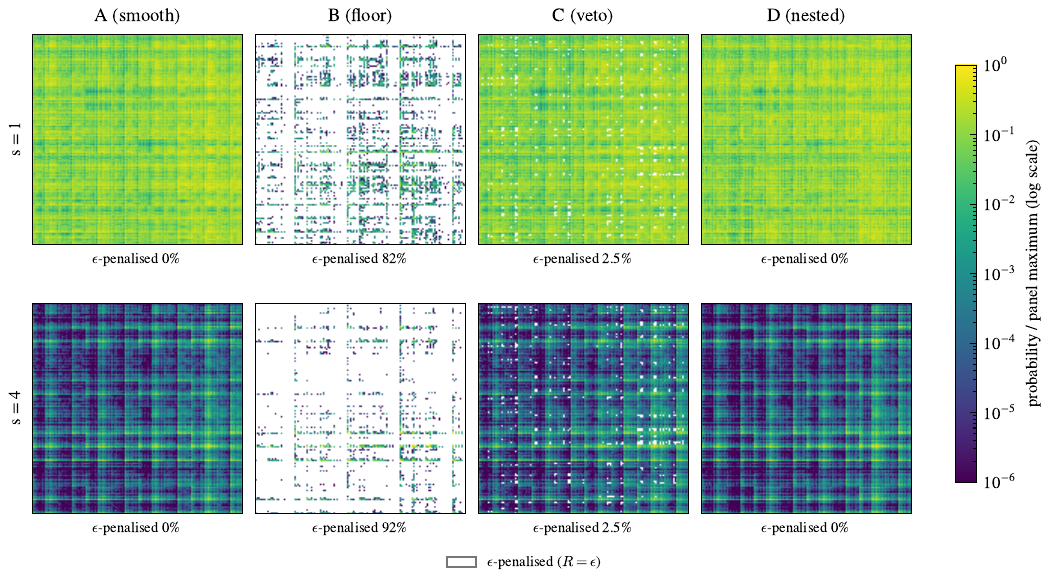}
  \caption{What each case asks for: exact targets \(p^*_c\) at the calibrated operating condition on the deployment-scale sequence family at reference sparsity (top) and in the sparse regime (bottom). White marks \(\epsilon\)-penalised candidates with the share reported under each panel. The colour encodes probability relative to each panel's own maximum, i.e. it compares candidates inside a panel and does not serve comparison across panels. The four geometries differ in kind: A and D stay dense, C's vetoes scatter isolated penalised points through an otherwise unpenalised space, and B's floor penalises most of the space, with sparsity taking most of what is left.}
  \label{fig:targets}
\end{figure*}

The four cases are constraint \emph{geometries}. They are fixed before any
experiment runs, and the risk parameters vary inside each of them. Their effect on the target is illustrated in
Figure~\ref{fig:targets}. Each case abstracts the requirements that arise when
a candidate is scored against multiple conformations of a target
(Table~\ref{tab:cases}). We take only that abstraction from the biology, and
use no further target information in the experiments.

Case A serves as our control. It requires binding to every conformation, with
none privileged, which gives a robust aggregate with no discontinuity. The
conjunctive shape is borrowed from targets such as a viral protease, which is
active only when two identical copies are joined \citep{ferreira2021}, so a
molecule must act on each.

Case B adds a floor. Some targets present conformations to bind alongside
others to avoid. In the FXR/RXR heterodimer, agonists trap the AF-2 helix in a
coactivator-competent state while antagonists destabilise it, so a candidate
intended as an antagonist must avoid the conformations that permit coactivator
recruitment \citep{diazholguin2022}. A candidate that falls too far short on
the conformations it must bind is discarded irrespective of what it achieves on
the ones to avoid. The reward trades the aggregate over the conformations to
bind against the aggregate over those to avoid at rate \(\gamma\), and
\(\epsilon\)-penalises whatever falls below a floor \(f\) on the former, a
condition no single score can trigger on its own. The floor penalises in bulk,
with 82\% of the deployment-scale space \(\epsilon\)-penalised at the reference
sparsity and 92\% at \(s=4\).

Case C adds vetoes. Certain conformations must never be bound at all. An
intended inhibitor can instead activate its target by triggering a change of
conformation \citep{kholodenko2015}, and a veto rules out binding that
conformation. Any candidate whose score on some conformation \(k\) reaches
\(c_k - \delta_v\) is \(\epsilon\)-penalised. This scatters isolated
\(\epsilon\)-penalised points (2.5\%) through an otherwise unpenalised space.

Case D nests. When a target's conformations fall into natural groups, such as
the repeated subunits of a symmetric protein whose binding sites interact
\citep{maltarollo2022}, scores are pooled within a group before the groups
combine. Here, an \emph{origin} is one such group. Nesting allows a separate radius per
origin, \(\rho_o\), under an outer tail level \(\beta_{\text{out}}\) that runs
from compensatory (\(\beta_{\text{out}} = 1\)) to conjunctive
(\(\beta_{\text{out}} \to 0\)).
This leaves the target dense.

To formalise the constraint geometries, we write \(\Psi(x)\) for the case's aggregation of the \(K\) scores of candidate
\(x\) into a single number, built from \(\Phi^{\pm}\) of Eq.~\eqref{eq:dro} as
Table~\ref{tab:cases} sets out per case. The reward and the target it induces
are
\begin{equation}
  R(x) =
  \begin{cases}
    \epsilon, & x \text{ penalised},\\
    \max(w_g\, g(x) + w_s \Psi(x),\, \epsilon), & \text{else},
  \end{cases}
  \label{eq:reward}
\end{equation}
\begin{equation}
  p^*_c(x) = \frac{R(x)^{\beta_t}}{\sum_{x' \in \mathcal{X}} R(x')^{\beta_t}},
  \label{eq:target}
\end{equation}
with \(w_s = 1 - w_g\) and \(\epsilon = 10^{-4}\). A candidate is
\emph{\(\epsilon\)-penalised} when it fails the floor (Case B) or a veto
(Case C); it keeps reward \(\epsilon\), so it keeps target mass and stays in the
support (Appendix~\ref{app:epsmass}). The condition vector \(c\)
collects the target temperature \(\beta_t\) (log-uniform on \((0.5, 8]\)), the
auxiliary weight \(w_g \in (0.1, 0.9)\), and the case's risk block. It varies
continuously, but the family it indexes does not. The floor is a contour
discontinuity in the aggregate, and the vetoes create \(\epsilon\)-penalised regions in
\(\mathcal{X}\). Therefore \(p^*_c\) jumps wherever a threshold crosses a
candidate. Table~\ref{tab:glossary} in Appendix~\ref{app:supp} collects every control parameter, with
its range and default. The smallest nominal weight of a set bounds its tail level from below. We
report the cells this rules out and never drop them silently. Because the
space is enumerable, the whole construction runs as one exact, training-free
computation vectorised over all of \(\mathcal{X}\) (Algorithm~\ref{alg:target},
Appendix~\ref{app:supp}). Every learned sampler in \S\ref{sec:results} is measured against this object.

\subsection{The conditional GFlowNet}
\label{sec:model}

A GFlowNet builds a candidate in steps. Its forward policy
\(P_F(\cdot \mid c)\) chooses each step, its backward policy \(P_B\) says how a
finished candidate could have been reached, and a learned scalar
\(\log Z(c)\) estimates the log partition function. Here candidates are
built coordinate by coordinate in a fixed canonical order with
\(P_B \equiv 1\), so each candidate has exactly one trajectory and the
terminating distribution \(p_\theta(\cdot \mid c)\), with \(\theta\) the
network's parameters, is a product of \(d\) softmaxes that can be enumerated. The
licence is the one \citet{shen2025} invoke in the 3D setting we ultimately
target: autoregressive assembly makes the backward path deterministic.

The policy is a residual multilayer perceptron (MLP) trunk (width 256, depth 4) over summed
position-tagged coordinate embeddings. It is modulated by Feature-wise Linear
Modulation (FiLM, \citealp{perez2018}) via a condition encoder (dimension 128),
with a learned head \(\log Z(c)\). Trajectory Balance (TB) collapses to the
per-point residual
\begin{equation}
  \delta(x, c) = \beta_t \log R(x) - \bigl[\log Z(c) + \log P_F(x \mid c)\bigr],
  \label{eq:tb}
\end{equation}
minimised as \(\mathbb{E}[\delta^2]\) over batches mixing on-policy samples with
uniform draws over \(\mathcal{X}\) (50/50). Every candidate therefore has support under the training distribution, and we
report what that choice costs as an ablation.
Conditions come from a pool of 256 with precomputed log-rewards. A training-only
clamp at \(-25\) on \(\beta_t \log R\) keeps \(\epsilon\)-region targets from
dominating residuals; evaluation targets are never clamped. The clamp binds
on \(\epsilon\)-penalised candidates when \(\beta_t \log \epsilon < -25\), that
is for \(\beta_t > 2.71\), which covers 39\% of the condition pool. Runs use 4,000
steps (12,000 for the learnability measurement of \S\ref{sec:learnability}) with
batches of 8 conditions \(\times\) 64 points and learning rates \(10^{-3}\)
(trunk) and \(10^{-2}\) (\(\log Z\)). Two controls accompany the main runs:
SubTrajectory Balance (SubTB) \citep{madan2022} with geometric segment weight
\(\lambda=0.9\), and an \emph{exact-KL oracle}: the same network trained by
cross-entropy against the enumerated target, which shows what training reaches
when the exact target is available. Remaining settings are in Appendix~\ref{app:config}.

\subsection{Evaluation protocol}
\label{sec:protocol}

We score amortisation by the exact mean \(L_1\) to \(p^*_c\) over a held-out
grid of conditions, fixed before training and listed in
Appendix~\ref{app:protocols}. The trainer rejects any condition it draws whose
\(L_\infty\) distance to a held-out condition falls below 0.05, measured in
the feature encoding the network receives, where each coordinate of \(c\) maps
onto roughly \([-1,1]\) over its range and \(\beta_t\) and the radii enter on
a log scale. Nothing the student trains on therefore comes closer than 0.05 to
anything it is scored on.

We read every error against a \emph{finite-sample reference}: the mean
\(L_1\) between \(p^*_c\) and an \(n\)-sample draw from \(p^*_c\) itself, at
\(n = 10\,|\mathcal{X}|\), which is 10,240 on the calibration family and
655,360 on the sequence family, averaged over 20 repetitions. The reference is
a declared scale and not a lower bound, since a perfect sampler has exact
\(L_1\) zero. What it measures is how far sampling noise alone moves an
estimate at the budget a deployment without an enumerable \(\mathcal{X}\)
actually sees. It falls as \(1/\sqrt{n}\), so a ratio to it grows as
\(\sqrt{n}\), and Appendix~\ref{app:refsweep} gives those ratios over a range
of \(n\), across which the ordering of the four cases does not change. Our
endpoint, fixed before the runs, is \(L_1 \leq 3\times\) the reference over 4
worlds \(\times\) 3 seeds.

Two other quantities appear in \S\ref{sec:results}. Total variation between
the risk-on and risk-off targets measures how far the risk parameters move the
target. \emph{Joint-satisfaction mass} measures utility: the probability the
target places on candidates meeting the world's challenge level, which is the
threshold a candidate must reach to count as successful. Under sparsity we
also report \emph{mode coverage}, the expected fraction of the mode set a
fresh budget of independent draws from the trained policy would find. A
\emph{mode} is a candidate satisfying the world's requirement at its challenge
level, and the size of that set is the denominator, reported per world. We
compute coverage from the trained policy, and not from whatever training
happened to visit, at reward-query budgets matched across regimes, including
the queries the teacher spends on itself. Appendix~\ref{app:protocols} defines
the challenge level, the mode set and the effective number of candidates in
full.

Worlds are the unit of replication, with seeds nested inside them. We report
cluster bootstrap confidence intervals \citep{field2007} and effect sizes
throughout, use paired world-level sign-flip tests where the design is paired
\citep{ernst2004}, use two one-sided tests for equivalence
\citep{schuirmann1987}, and apply the Holm correction across tests on the same
data \citep{holm1979}. A paired test on four worlds cannot resolve below
\(p = 0.125\), so at that replication we give intervals and effect sizes and
make no significance claim.

\section{Results}
\label{sec:results}

\subsection{How the risk parameters change the target}
\label{sec:aggregation}

Conditioning on \((\beta, \rho)\) is worth its training cost only if the
targets they index differ from one another. For each case we compute the exact
target on an \(11 \times 9\) grid over \((\beta, \rho)\) for eight calibration
worlds. We call one point of that grid a \emph{cell}, and we score every cell
by joint-satisfaction mass, the probability
the target places on candidates that meet the world's challenge level.
Comparisons are paired within worlds, and each setting contributes its best
admissible cell in a world before we average.

\begin{table*}[tb]
  \centering
  \caption{Joint-satisfaction mass at the challenge level, with the change in
  the effective number of distinct candidates in the last column (best
  admissible cell per setting, paired within worlds, mean over 8 calibration
  worlds). Weight distrust on its own improves on the weighted mean in 8 of 8
  worlds in every case. The last column compares the most conservative setting
  with the weighted mean. Because satisfaction mass rises monotonically in
  both parameters, an argmax over all risk cells lands on the grid corner
  nearest the most conservative setting, in 8 of 8 worlds on the smooth case
  and 7 of 8 on the veto case, so such an argmax does not identify a separate
  intermediate setting.}
  \label{tab:knobs}
  \small
  \begin{tabular*}{\textwidth}{@{\extracolsep{\fill}}l*{5}{c}@{}}
    \toprule
    Case & weighted mean & \(\rho\) only & \(\beta\) only & most conservative & \(\Delta\) candidates \\
    \midrule
    A smooth & 0.025 & 0.052 & 0.057 & 0.063 & \(-1.8\%\) \\
    B floor  & 0.058 & 0.511 & 0.547 & 0.601 & \(-32.5\%\) \\
    C veto   & 0.028 & 0.053 & 0.057 & 0.061 & \(+0.2\%\) \\
    D nested & 0.025 & 0.056 & 0.057 & 0.069 & \(-3.1\%\) \\
    \bottomrule
  \end{tabular*}
\end{table*}

Table~\ref{tab:knobs} shows three things. Both parameters change the target on
their own: weight distrust alone improves on the weighted mean in all 32
case-world pairs, and on the floor case it supplies most of the available
gain, 0.511 of 0.601. The floor case moves furthest, from 0.058 to 0.601,
while the smooth and veto cases gain a factor of two to two and a half. The
large gain costs diversity: the floor case gives up 32.5\% of its effective
candidates, and the other three cases stay within 3\% of where they started.

\begin{figure*}[tb]
  \centering
  \includegraphics[width=\textwidth]{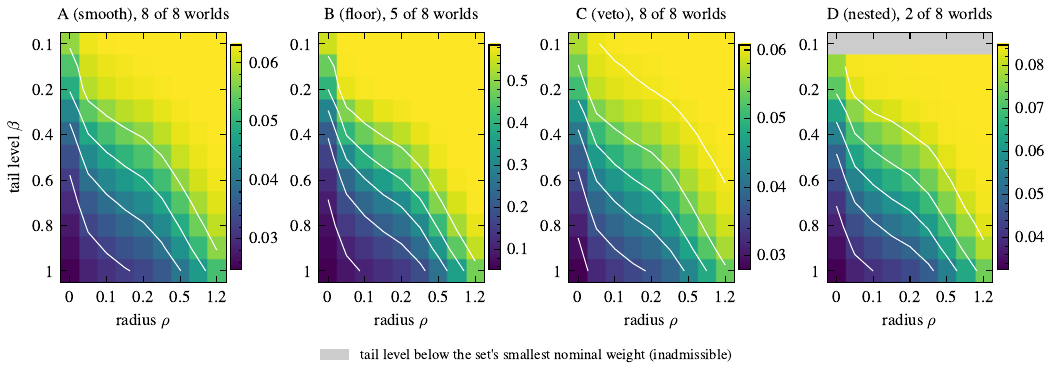}
  \caption{Joint-satisfaction mass across the \((\beta, \rho)\) plane, one
  panel per case, at each world's challenge level. Risk increases upwards
  (\(\beta\) from 1 to 0.1) and to the right (\(\rho\) from 0 to 1.2), so the
  weighted mean sits at the bottom left. A tail level cannot fall below the
  smallest nominal weight of the set it acts on, and that bound is higher for
  the nested case, from 0.116 to 0.194 across its worlds, than for the other
  three, which reach 0.010. Cells that bound rules out are grey: the row at
  the top of the nested panel is \(\beta = 0.1\), inadmissible in all 8 of
  its worlds. The same bound leaves
  a common grid of 5 of 8 worlds on the floor case and 2 of 8 on the nested
  case, and each panel states its count. White contours mark
  iso-satisfaction levels, and their diagonal run is the substitution the
  text describes.}
  \label{fig:surface}
\end{figure*}

Figure~\ref{fig:surface} gives the surface itself, which is the object a
practitioner works with. In deployment nobody selects a cell: \(\rho\) follows
from the quality of the population estimates, and \(\beta\) from how strictly
every conformation must be cleared. What the surface shows is that the two
settings substitute for each other over a long range. On the five floor worlds
where every cell of the plane is admissible, holding \(\beta = 1\) and raising
\(\rho\) alone moves satisfaction from 0.053 to 0.473, and holding
\(\rho = 0\) and lowering \(\beta\) alone moves it from 0.053 to 0.544. The
iso-satisfaction contours run diagonally, which is what that substitution
looks like on the plane.

\paragraph{Comparison with other conjunctive rules.}
Satisfaction mass alone cannot show that ours is the right conjunctive rule,
since any rule that concentrates mass on candidates passing every conformation
scores well on it. We therefore compare it with three alternatives in which
only the aggregation \(\Psi\) changes (Table~\ref{tab:baselines}). An
aggregation over an ensemble has to meet two requirements at once. It must
refuse to let a strong score on one conformation offset a failure on another,
and it must take the conformational populations into account, since these are
not uniform. Each alternative meets one of the two. The hard minimum refuses
the offset but ignores the populations, so the worst conformation decides
however rarely it is occupied. The weighted geometric mean uses the
populations but allows the offset, because it is a product. Derringer-Suich
desirability \citep{derringer1980} also uses the populations and also allows
the offset, and its ramp sends every score below the lower quantile to 0 and
every score above the upper quantile to 1, so candidates that clear a
conformation by different margins receive the same value, as do candidates
that fail it by different margins. Our rule is the only one of the four that
both refuses the offset and weights by population, and its tail level
\(\beta\) sets how much of the ensemble a candidate must clear.

\begin{table*}[tb]
  \centering
  \caption{The composed rule against three other conjunctive rules on the
  calibration family (8 worlds, \(\beta_t = 4\), \(w_g = 0.3\); the composed
  rule at \(\beta = 0.25\), \(\rho = 0.5\)). Only the aggregation \(\Psi\)
  changes: worlds, auxiliary objective, temperature and exclusion rules are
  shared, and the floor of Case B sits at the same quantile of each rule's own
  aggregate, since the rules score on different scales. Each entry gives
  joint-satisfaction mass and the effective number of distinct candidates,
  averaged over four thresholds per world: 0.25, 0.35, 0.45 and the world's
  challenge level. The last column gives the share of \(\mathcal{X}\) at
  reward \(\epsilon\) for the four rules in column order. Desirability maps
  each score through a ramp between its 0.50 and 0.95 quantiles.}
  \label{tab:baselines}
  \small
  \begin{tabular*}{\textwidth}{@{\extracolsep{\fill}}lccccc@{}}
    \toprule
    Case & composed & hard minimum & geometric mean & desirability & share at \(\epsilon\) (\%) \\
    \midrule
    A smooth & 0.370 / 145 & 0.370 / 145 & 0.256 / 157 & 0.365 / 62 & 0 / 0 / 0 / 0 \\
    B floor  & 0.752 / 10  & 0.752 / 10  & 0.233 / 42  & 0.254 / 22 & 89.6 / 89.6 / 38.7 / 39.4 \\
    C veto   & 0.433 / 153 & 0.434 / 153 & 0.317 / 165 & 0.502 / 52 & 15.1 / 15.1 / 15.1 / 15.1 \\
    D nested & 0.102 / 29  & 0.111 / 29  & 0.052 / 32  & 0.051 / 18 & 0 / 0 / 0 / 0 \\
    \bottomrule
  \end{tabular*}
\end{table*}

We report this comparison as a characterisation. The composed rule and the
hard minimum agree to three digits on the smooth, floor and veto cases, and
the hard minimum is slightly ahead on the nested case, because the two differ
only in whether the populations are weighted and satisfaction mass at this
setting does not register that difference. The weighted geometric mean has
lower satisfaction mass in every case, at a similar or larger number of
candidates, which is what allowing the offset produces. Desirability has the
highest satisfaction mass on the veto case, 0.502 against 0.433, and it holds
52 effective candidates there against our 153; on the smooth, veto and nested
cases it keeps between a third and two thirds of the candidates we keep. The
cause is saturation: every candidate above the upper quantile receives the
same top value, so the target concentrates on that slice. The same comparison
on the deployment-scale family repeats the pattern and adds one more loss,
where desirability takes the smooth case, 0.487 against 0.425, again at about
a third of the candidates. What separates our rule from the alternatives is
therefore mainly the number of distinct candidates it keeps while refusing the
offset.

The last column names a cost of that refusal. On the veto case all four rules
give reward \(\epsilon\) to the same 15.1\% of the space, since a veto depends
on the scores and not on the aggregation, and on the smooth and nested cases
no rule gives \(\epsilon\) to any candidate. The floor case separates them.
The floor itself fails the same 35.1\% of the space under every rule, because
it sits at the same quantile of each rule's own aggregate. Beyond that, a rule
that refuses the offset drives the promoted aggregate minus \(\gamma\) times
the suppressed one below \(\epsilon\) for most of what remains, so our rule
and the hard minimum give reward \(\epsilon\) to 89.6\% of the space, against
38.7\% for the geometric mean and 39.4\% for desirability. A region of
identical reward that large is the condition under which \S\ref{sec:sparsity}
finds a degenerate optimum for on-policy training.

\paragraph{Placing the radius where the scores are least reliable.}
Case B has two score sets with separate radii, and we draw its worlds with a
different reliability for each set, so we can ask whether the larger
radius belongs on the less reliable one. Over 8 worlds and 6 matched pairs per
world, placing it there improves stressed satisfaction at the 5th percentile
by 0.012 (paired sign-flip \(p = 0.030\), \(d = 0.76\)). Two confirmatory
tests were run on these data, and under the Holm correction of
\S\ref{sec:protocol} the comparison gives \(p = 0.060\), so we report the
effect and its size without a significance claim. The effect also does not
grow with the degree of asymmetry (Spearman \(r_s = -0.17\)), which a
mechanism acting through reliability would predict, so we read it as
suggestive only.

The nested case is the exception to the pattern in Table~\ref{tab:knobs}. No
interior cell of its plane exceeds the most conservative setting in any of the
8 worlds, 0.063 against 0.069, so on that case the risk parameters are worth
setting but the interior of the plane offers nothing over its corner.

\paragraph{Summary.}
Both parameters change the target in every world, and the composed setting
reaches satisfaction levels that neither parameter reaches alone. The gain is
a movement along a surface that trades satisfaction against diversity, and on
the floor case that trade costs about a third of the distinct candidates.
Against three other conjunctive rules at matched exclusions, what distinguishes
the composed rule is mainly how many distinct candidates it keeps while
refusing to let one score offset another.

\subsection{Amortising the family in one network}
\label{sec:learnability}

We now ask whether one network can serve the whole family. We train a single
conditional student per world and seed and score it by exact \(L_1\) on the
held-out condition grid. Beside it we train an exact-KL oracle, the same
architecture fitted by cross-entropy to the enumerated target. The oracle
shows what this architecture and this optimiser reach when the target is
handed to them, so a student that trails an oracle which meets the endpoint
has an optimisation shortfall. An oracle that misses the endpoint itself
leaves the question open at that budget, since cross-entropy training may
simply not have converged. Errors are read against the finite-sample
reference of \S\ref{sec:protocol}. The floor case carries the largest error throughout, because its target \(\epsilon\)-penalises most of the space, as Figure~\ref{fig:targets} shows.

\begin{table}[tb]
  \centering
  \small
  \caption{Learnability at deployment scale (sequence family, 65{,}536 states
  at depth 8, 12{,}000 steps, 12 runs per case). Exact held-out \(L_1\) as a
  multiple of the finite-sample reference. This is the headline measurement;
  Table~\ref{tab:learnability} gives the same protocol on the smaller
  calibration family.}
  \label{tab:learnability-deploy}
  \begin{tabular}{@{}lcccc@{}}
    \toprule
     & smooth & floor & veto & nested \\
    \midrule
    student, \(\times\) reference & 0.39 & 2.85 & 1.51 & 0.53 \\
    oracle, \(\times\) reference  & 0.18 & 0.95 & 0.28 & 0.33 \\
    \bottomrule
  \end{tabular}
\end{table}

\begin{table*}[tb]
  \centering
  \caption{Learnability on the calibration family (12,000 steps, 4 worlds \(\times\) 3 seeds). Exact held-out \(L_1\) against the finite-sample reference. Student and oracle are trained on the same four worlds, so the gap between them is a paired quantity; we give it with cluster bootstrap intervals and paired effect sizes. At four worlds a paired sign-flip test cannot resolve below \(p = 0.125\), so we make no significance claim.}
  \label{tab:learnability}
  \small
  \begin{tabular*}{\textwidth}{@{\extracolsep{\fill}}l*{4}{c}p{5.0cm}@{}}
    \toprule
    Case & \(L_1\) & ref. & ratio & oracle & where the error sits \\
    \midrule
    A smooth & 0.082 & 0.223 & \textbf{0.37\(\times\)} & 0.17\(\times\) & both below the reference \\
    \addlinespace[2pt]
    B floor  & 0.398 & 0.147 & \textbf{2.7\(\times\)} & 1.18\(\times\) & oracle also above it \\
    \addlinespace[2pt]
    C veto   & 0.338 & 0.198 & \textbf{1.7\(\times\)} & 0.44\(\times\) & student only \\
    \addlinespace[2pt]
    D nested & 0.133 & 0.221 & \textbf{0.60\(\times\)} & 0.45\(\times\) & both below the reference \\
    \bottomrule
  \end{tabular*}
\end{table*}

All four cases meet the endpoint of \(3\times\) the reference at deployment
scale (Table~\ref{tab:learnability-deploy}) and on the calibration family
(Table~\ref{tab:learnability}). The smooth and nested cases sit below the
reference on both families, so conditioning on the risk parameters costs
nothing there. The veto case separates the two sources of error cleanly: its
oracle reaches \(0.44\times\) the reference while its student sits at
\(1.7\times\), so the whole gap belongs to the student's optimisation. The
student-oracle gaps are 0.044 [0.034, 0.056] on the smooth case, 0.225
[0.208, 0.242] on the floor case, 0.252 [0.236, 0.267] on the veto case and
0.033 [0.025, 0.042] on the nested case, with paired effect sizes of 5.0,
11.7, 22.7 and 6.7.

The floor case is the one whose oracle also stays above the reference, at
\(1.18\times\). Two controls bear on how to read that. The same oracle on the
same case reaches \(0.95\times\) at deployment scale, on 65,536 states, which
an architecture short of capacity would not do on the larger problem. Widening
the trunk moves it in the right direction and does not cross the line: at
widths 128, 256 and 512 the floor case's oracle reaches \(1.26\times\),
\(1.18\times\) and \(1.12\times\). We therefore read the calibration-scale
miss as training that has not converged at this budget, and we do not claim a
limit of the architecture.

One asymmetry inside this comparison deserves its own control. The clamp of
\S\ref{sec:model} applies to the trajectory-balance student alone, while the
oracle's targets are exact and unclamped. Removing it makes the student worse
on both hard cases, from \(2.70\times\) to \(4.39\times\) on the floor case
and from \(1.70\times\) to \(3.08\times\) on the veto case, while the oracle
stays where it was, \(1.19\times\) against \(1.18\times\) and \(0.43\times\)
against \(0.44\times\). The clamp therefore narrows the gap we attribute to
the student and does not create it.

Amortisation costs something. Per-condition experts reach \(1.4\times\) to
\(3.4\times\) lower error than the conditional student at matched compute, and
each of them serves one condition, so the saving from one network is paid for
in held-out error.

\paragraph{Summary.}
One conditional network learns the whole family inside the endpoint, with two
of the four cases below the reference. On the veto case the oracle places the
residual error in the student's optimisation. On the floor case the oracle
misses the reference at calibration scale, meets it at deployment scale, and
improves with width, so that residual is training at this budget.

\subsection{Exploration under sparse rewards}\label{sec:sparsity}

The results so far hold where satisfying candidates are common. Raising the
sparsity exponent \(s\) removes that assumption. Figure~\ref{fig:targets} 
(bottom row) shows what that does to each target. On the deployment-scale
sequence family we train an unconditional student at the calibrated condition
under five regimes that share every setting except the source of the
off-policy half of each batch: none (on-policy), uniform over \(\mathcal{X}\)
(mix), reward-prioritised replay, an adaptive teacher \citep{kim2024}, and a
two-buffer contrastive replay scheme \citep{kim2026}. The five fall into two
groups. Injection (mix, teacher) supplies candidates the policy would not
visit on its own. Reweighting (replay, contrastive replay) redistributes over
candidates it has already produced.

\begin{table*}[tb]
  \centering
  \caption{Mode coverage of the \emph{trained policy} at sparsity \(s = 4\):
  the expected fraction of the mode set a fresh evaluation budget of
  independent draws would find, averaged over 4 worlds \(\times\) 3 seeds at
  reward-query budgets matched across regimes, including the teacher's. Modes
  the reward itself penalises are left out of the set. The last column gives
  the same quantity for the exact target, which is what a perfect sampler
  reaches at this budget, so it is the reference each row is read against and
  1 is not attainable. At \(s = 1\) every regime is within 0.01 of the
  reference on the smooth, veto and nested cases; on the floor case the five
  span 0.820 to 0.966 against a reference of 0.988.}
  \label{tab:arms}
  \small
  \begin{tabular*}{\textwidth}{@{\extracolsep{\fill}}l*{6}{c}@{}}
    \toprule
    & \multicolumn{2}{c}{injection} & \multicolumn{2}{c}{reweighting} & & \\
    \cmidrule(lr){2-3}\cmidrule(lr){4-5}
    Case & mix & teacher & replay & contr. & on-pol. & target \\
    \midrule
    A smooth & 0.904 & 0.906 & 0.857 & 0.806 & 0.836 & 0.950 \\
    B floor  & 0.518 & 0.518 & \textbf{0.725} & 0.616 & 0.305 & 0.780 \\
    C veto   & 0.956 & 0.939 & 0.936 & 0.895 & 0.913 & 0.981 \\
    D nested & 0.557 & 0.563 & 0.518 & 0.499 & 0.510 & 0.646 \\
    \bottomrule
  \end{tabular*}
\end{table*}

Sparsity changes what binds, and not along the line the two groups would
suggest. At \(s = 4\) the on-policy regime falls far below the reference on
the floor case, 0.305 against 0.780, while every off-policy regime stays
between 0.518 and 0.725 (Table~\ref{tab:arms}). The first division is
therefore whether any off-policy source is present at all.

Meeting the modes during training is not what decides the rest. On the floor
case at \(s = 4\) uniform mixing meets 99.7\% of the mode set while training
and reaches 80\% of it after 192,000 samples, and its trained policy still
recovers only 0.518. Replay meets 38.5\% of the set on average and its policy
recovers 0.725. The two leave almost the same total mass on modes, 0.178
against 0.173, and differ in how widely they spread it. What decides the
difference is where the training signal lands. At \(s = 4\) on this case 92\%
of the space sits at reward \(\epsilon\), one identical value, so a uniform
off-policy half spends most of its budget teaching the policy a constant,
while a reward-prioritised half spends almost all of it where the modes differ
from one another. At a matched reward-query budget it is that allocation the
final coverage follows, and the exact \(L_1\) agrees: 0.291 for replay against
0.505 for mixing.

\begin{figure*}[tb]
  \centering
  \includegraphics[width=0.8\textwidth]{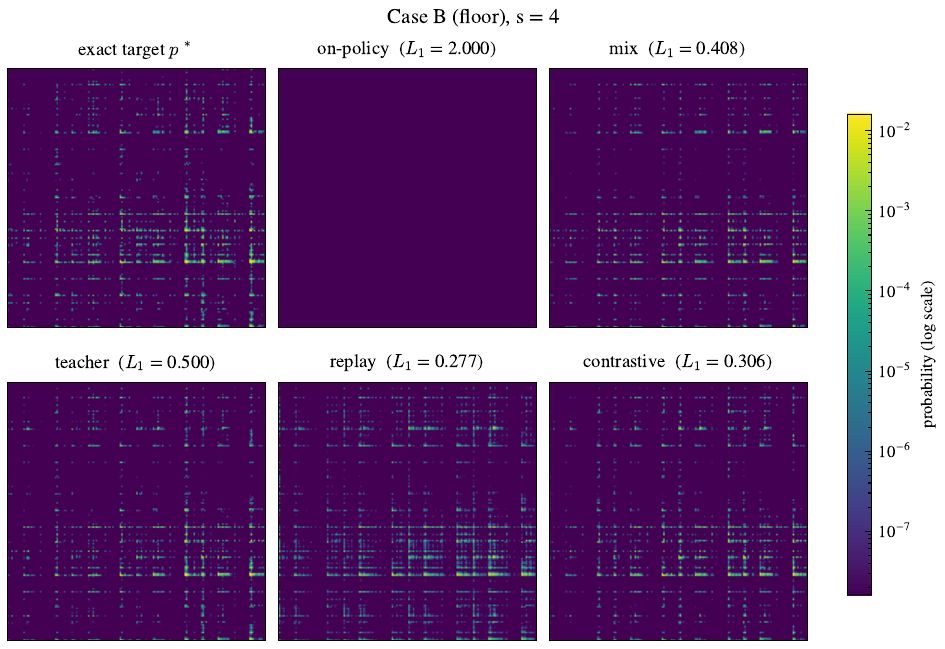}
  \caption{What each regime reaches on the floor case at \(s=4\), for one
  world and seed. The on-policy panel is empty because its \(L_1\) reaches
  2.000 to the three decimals shown, falling short of 2 by about
  \(2 \times 10^{-16}\). Its support is therefore almost disjoint from the
  target's, as far apart as two distributions can be while both stay strictly
  positive, without being disjoint, which a strictly positive reward rules
  out. This world is not the typical one: 5 of the 12 on-policy runs
  enter the degenerate optimum and 7 do not, giving a mean \(L_1\) of 1.033.
  Averaged over worlds and seeds, replay takes both the better exact \(L_1\)
  on this case, 0.291 against 0.505 for mixing, and the higher coverage of the
  trained policy (Table~\ref{tab:arms}).}
  \label{fig:collapse}
\end{figure*}

The mechanism is visible in the exact data. Every candidate inside the
\(\epsilon\)-plateau has the same reward, so the trajectory-balance residual
over that region is flat, and a policy that puts all its mass on one
\(\epsilon\)-penalised candidate and sets
\(\log Z = \beta_t \log \epsilon\) drives its empirical loss on its own
samples to zero. That configuration is a degenerate optimum, and with 92\% of
the space inside the plateau an on-policy learner begins there.
Figure~\ref{fig:collapse} shows a world in which it never leaves. The
injecting regimes never reach it, because a single live candidate in the batch
pushes \(\log Z\) back up. The reweighting regimes cannot recover a candidate
they never produced, and they concentrate near the edge of the plateau:
contrastive replay draws 95\% of its samples from candidates one coordinate
away from a penalised one, the highest of the five, against 68\% for uniform
mixing. A rising share of such boundary samples together with a coverage that
stops improving is what makes the degenerate optimum visible during training,
without any enumeration (Figure~\ref{fig:traces},
Appendix~\ref{app:densities}).

\paragraph{Summary.}
Where a large exactly flat region of reward exists, on-policy training settles
into a degenerate optimum and any off-policy source prevents it. Past that
point, what limits the trained sampler is how much of its training signal
lands on the live region, and not how much of the space it visited: the regime
that meets almost every mode recovers fewer of them than the one that meets
about a third. The finding is about that geometry, which our floor
construction creates and sparsity deepens, and it does not extend to sparse
rewards in general. On the smooth, veto and nested cases, which stay dense
under the same exponent, every regime including on-policy stays within 0.15 of
the reference.

\subsection{Cost and failure modes}
\label{sec:price}

Trainability is not uniform over the design (Appendix~\ref{app:prices}). Two
axes cost something that a larger budget recovers. Resolution costs the veto
case: the boundary of the penalised region grows with the space and has to be
visited before it can be fitted, so the error climbs from \(1.4\times\) to
\(2.1\times\) the reference between \(H = 16\) and \(H = 64\), while the
smooth and nested cases hold their error across the same range. Depth costs
the floor case, and only because the budget runs out first. At \(d = 3\) and
\(d = 4\) the student passes the endpoint while its error is still falling,
and tripling the budget brings it back to \(2.9\)--\(3.0\times\).

Two axes resist a larger budget. The peaked weight family (Dirichlet
concentration 0.3) leaves the floor case at \(3.2\times\), where tripling the
budget does not move it. We report that as an open failure. Conditioning on
the score-noise margin \(\sigma\) also breaks the endpoint on both hard cases,
and here the oracle locates the problem: the floor case's oracle under
\(\sigma\) is itself beyond \(3\times\) the reference, so training does not
reach the \(\sigma\)-indexed family even when the target is handed to it. A
corruption study points the same way, since pre-shifting the scores by
\(\sigma\) roughly doubles the error to the true target, from
\(\mathrm{TV} = 0.20\) to 0.38 at 10\% corruption.

The submitted version of this paper advised spending that uncertainty on the
ambiguity radius instead. A matched comparison does not support the advice, so
we withdraw it. On a single grid scored by the realised plain weighted mean
under the truth, so that no parameter is judged by its own limiting case, we
applied score error and weight error separately and together. Across all 48
combinations of case, error model and error size, the cell that maximises the
stressed criterion has \(\rho = 0\), with the score-robustness margin at its
largest grid value throughout. What moves under score error is the tail level:
on the floor case it rises from 0.6 at \(\sigma = 0\) to 1.0 at
\(\sigma = 0.2\), and on the nested case from 0.3 to 0.5, while weight error
alone leaves it where it was. Neither kind of error is best handled by the
radius.

As the auxiliary objective takes more of the reward, the risk parameters stop
changing the target: at weight 0.9 on the smooth and veto cases and 0.85 on
the nested case. The floor case keeps responding throughout, staying
\(\mathrm{TV} \geq 0.26\) apart even at 0.99. At \(s = 4\) the nested
aggregation falls back to its flat form in 199 or 200 of 200 draws.

\paragraph{Summary.}
The reward is trainable, at a cost that two of the four axes recover with a
longer budget. Peaked weights are an open failure, and score noise defeats
training even when the target is exact. The radius does not absorb either kind
of error.

\section{Limitations}
\label{sec:limitations}

Because the candidate space here is small enough to list in full, we can
measure the error exactly against the distribution a sampler ought to produce.
A real chemical space cannot be listed. The reward and the training method
transfer to a real target unchanged, but this measurement does not, because
there is no exact target to compare a sampler with.

We set the four difficulty axes by construction and discard any world that
turns out to be degenerate. A real project has neither option, since its
scores come from whatever scoring function and conformer ensemble are in use,
and its target is assigned and cannot be set aside for being too easy or too
hard. Appendix~\ref{app:gate} reports how often the gate rejects a world, and
it is selective on the nested case, accepting roughly one draw in four. Our
numbers therefore describe a population of problems we selected, and they do
not predict the cost of any particular target.

The oracle shows what cross-entropy training reaches when the target is known
at a fixed budget. An oracle that meets the endpoint shows the architecture
can represent the target; an oracle that misses it leaves the question open,
because training may not have converged. We draw no conclusion about
representational capacity from an oracle failure.

Our largest space holds 65,536 candidates at depth 8. The trends we report
along size and depth are consistent inside that range, and we do not know
whether they hold in spaces orders of magnitude larger.

Two results rest on narrow evidence. The response surface of
\S\ref{sec:aggregation} uses the worlds in which every cell of the plane is
admissible, which is 5 of 8 worlds on the floor case and 2 of 8 on the nested
case. The placement result for the radius does not survive the Holm
correction, and we report it as suggestive.

Finally, the scores are analytic. Nothing here compares the composed reward
with docking, and the satisfaction numbers are properties of our own score
fields.

\section{Conclusion}
\label{sec:conclusion}

One GFlowNet learns the whole family of targets, staying within
\(0.39\)--\(2.85\) times the error that sampling noise alone contributes at
our evaluation budget. Where satisfying candidates are rare, the limit on the
sampler is which candidates its training signal reaches, and a reward that is
exactly flat over most of the space moves that limit well before the fit to
any one target does.

The construction maps onto structure-based drug design. Docking scores against
each conformation become the \(K\) scores, and the conformational populations
become the nominal weights. The tail level \(\beta\) sets how strictly a
candidate must satisfy all of them at once, and the radius \(\rho\) sets how
far those estimated shares are trusted. The floor and the veto express what a
ranking leaves out, that some conformations must be cleared and others never
bound.

The scores here are analytic, so what transfers is the construction and the
diagnostics. Whether the composed reward helps on a real target is an open
question, and answering it needs docking scores against a real conformational
ensemble, where the exact target this paper measures against no longer exists.

\section*{Code and data}
The reward construction, the world generator, the training and evaluation
scripts and the analysis behind every table are at
\url{https://github.com/carineribeirost/risk-sensitive-gfn}, together with a
script that reproduces the runs reported here.
No external data is used: every world the paper reports is regenerated by
that code from the seeds recorded with each run.

\printbibliography

\clearpage
\appendix
\onecolumn
\section*{Appendix}
\addcontentsline{toc}{section}{Appendix}

\section{Order preservation under temperature and under the risk parameters}
\label{app:reorder}

Section~\ref{sec:intro} claims that the family \((\beta,\rho)\) indexes cannot
be reached by sweeping the temperature of a single target. The argument has two
halves, both checkable directly, and both are about the targets themselves
and not about any policy parameterisation.

First, temperature preserves the order of the target family. The target is
\(p^*_c \propto R(x)^{\beta_t}\), which is monotone in \(R\), so any two
candidates keep their relative order at every \(\beta_t > 0\). Sweeping the
temperature reweights the family but cannot reorder it.

Second, the risk block does not. Table~\ref{tab:reorder} evaluates \(\Phi^-\) at
nominal weights \((0.40, 0.35, 0.25)\) for \(A = (0.95, 0.90, 0.10)\), strong on
two signals and failing the third, and \(B = (0.62, 0.60, 0.58)\), uniformly
mediocre: at the Boltzmann pole \(A\) wins, and at any cell with a deepened tail
or a widened radius \(B\) does. One network has to serve both cells, and no logit scaling can invert a
ranking, so the risk parameters have to enter as conditioning.
\S\ref{sec:learnability} measures what that costs.

\begin{table}[htbp]
  \centering
  \small
  \caption{The ranking inverts inside the family. \(\Phi^-\) at nominal weights \((0.40,0.35,0.25)\) for \(A=(0.95,0.90,0.10)\) and \(B=(0.62,0.60,0.58)\), TV ball. The inversion is not specific to this geometry: any coherent risk measure with a non-trivial tail level reorders candidates whose score vectors differ in shape while sharing a mean.}
  \label{tab:reorder}
  \begin{tabular}{@{}lccc@{}}
    \toprule
    \((\beta, \rho)\) & \(\Phi^-(A)\) & \(\Phi^-(B)\) & winner \\
    \midrule
    \((1.00, 0.0)\) Boltzmann pole & 0.720 & 0.603 & \(A\) \\
    \((0.50, 0.0)\) tail only      & 0.500 & 0.590 & \(B\) \\
    \((0.25, 0.0)\) tail only      & 0.100 & 0.580 & \(B\) \\
    \((1.00, 0.3)\) distrust only  & 0.465 & 0.591 & \(B\) \\
    \((0.25, 0.3)\) interior       & 0.100 & 0.580 & \(B\) \\
    \((0.25, 0.5)\) interior       & 0.100 & 0.580 & \(B\) \\
    \bottomrule
  \end{tabular}
\end{table}

\clearpage
\section{Notation and the exact target}
\label{app:supp}

Table~\ref{tab:glossary} lists every control parameter of \S\ref{sec:targets}
with its range and default. It contains no learned parameter and no world
constant: every row is something a practitioner would set, which is what makes
the family one a single network can be conditioned on.
Algorithm~\ref{alg:target} states the target computation in full. Both are
training-free: \(p^*_c\) exists before any sampler does, and is
the object every \(L_1\) in \S\ref{sec:results} measures distance to. The
ambiguity radii are swept on geometry-specific grids, which are not comparable
across balls: \(0, 0.2, 0.5, 1.2\) nats for the KL ball, \(0, 0.05, 0.15, 0.35\)
of moved mass for total variation, and \(0, 0.3, 1, 3\) for the modified
\(\chi^2\) ball. All three duals were checked against brute force over 200
random instances per ball. Agreement reaches a floor of \(\approx 3 \cdot
10^{-8}\) across every geometry. This is an observed agreement level and not a derived error bound: the golden-section searches run 40 iterations, shrinking
the bracketed interval by \(\approx 7 \cdot 10^{-9}\), but we do not establish
a formal bound on the resulting error in the value. Residuals above that level come from the coarseness of the brute-force grid,
on two grounds the check tests directly: the implementation never exceeds the grid infimum, which it would if
a real gap existed, and the residuals collapse under refinement. At
\(\rho = 0\) the computation short-circuits to the plain CVaR and is exact by
construction.

\begin{table}[htbp]
  \centering
  \small
  \caption{Every control parameter in the reward family and what it does.
  Table~\ref{tab:notation} lists the remaining symbols. Each symbol here has
  one meaning throughout the paper.}
  \label{tab:glossary}
  \begin{tabular}{@{}p{1.5cm}p{8.0cm}p{5.4cm}@{}}
    \toprule
    Symbol & meaning & range / default \\
    \midrule
    \(\beta\) & CVaR tail level, per set: 1 compensatory, \(\to 0\) conjunctive & \((0,1]\), \(\geq\) set's smallest weight \\
    \addlinespace[2pt]
    \(\rho\) & ambiguity radius, per set: distrust of the stated weights & \([0, 1.6]\) (KL ball) \\
    \addlinespace[2pt]
    \(\beta_t\) & target temperature: sharpness of \(p^* \propto R^{\beta_t}\) & \((0.5, 8]\) log-uniform \\
    \addlinespace[2pt]
    \(a\), \(w\) & the \(K\) scores of a candidate and their nominal weights & \([0,1]^K\); Dirichlet(2.0) \\
    \addlinespace[2pt]
    \(g\), \(w_g\) & auxiliary objective (e.g., drug-likeness, \citealp{bickerton2012}) and its weight; \(w_s = 1 - w_g\) & \((0.1, 0.9)\) \\
    \addlinespace[2pt]
    \(\gamma\), \(f\) & trade-off rate for suppressed scores; floor threshold (Case B) & 1; 35\% quantile, per world \\
    \addlinespace[2pt]
    \(c_k\), \(\delta_v\) & veto threshold and margin (Case C) & 0.85; \([0, 0.1]\) \\
    \addlinespace[2pt]
    \(\rho_o\), \(\beta_{\text{out}}\) & a separate radius for each origin, and the outer tail level (Case D) & as \(\rho\), \(\beta\) above \\
    \addlinespace[2pt]
    \(s\) & sparsity exponent on objective fields (constraints exempt) & 1 (identity); 4 in sparse regime \\
    \addlinespace[2pt]
    \(\sigma\) & score-robustness margin: good \(-\sigma\), bad \(+\sigma\) & \([0, 0.2]\) \\
    \addlinespace[2pt]
    \(\epsilon\) & reward floor on penalised and non-positive states & \(10^{-4}\) \\
    \bottomrule
  \end{tabular}
\end{table}

\begin{table}[htbp]
  \centering
  \small
  \caption{The remaining notation, with the section that introduces each
  symbol. Every object has one name: the \(K\) scored things are
  \emph{conformations} and the functions scoring them are \emph{score
  fields}; an element of \(\mathcal{X}\) is a \emph{candidate}; a candidate
  the reward sends to \(\epsilon\) is \emph{\(\epsilon\)-penalised}.}
  \label{tab:notation}
  \begin{tabular}{@{}p{2.0cm}p{9.0cm}p{2.2cm}@{}}
    \toprule
    Symbol & meaning & introduced \\
    \midrule
    \(\mathcal{X}\), \(H\), \(d\) & candidate space \(\{0,\dots,H-1\}^d\), alphabet size, length & \S\ref{sec:testbed} \\
    \addlinespace[2pt]
    \(x\) & a candidate, one element of \(\mathcal{X}\) & \S\ref{sec:testbed} \\
    \addlinespace[2pt]
    \(K\), \(a_k\), \(a(x)\) & number of conformations, the score field of conformation \(k\), the vector of all \(K\) scores of \(x\) & \S\ref{sec:testbed} \\
    \addlinespace[2pt]
    \(S^0, S^+, S^-\) & the neutral, promoted and suppressed score sets & Table~\ref{tab:cases} \\
    \addlinespace[2pt]
    \(o\), \(O\), \(\pi\) & an origin (one group of conformations), their number, and the weights across them (Case D) & \S\ref{sec:targets} \\
    \addlinespace[2pt]
    \(\Phi^-\), \(\Phi^+\) & the DRO-CVaR functional on a set to be kept high or kept low, Eq.~\eqref{eq:dro} & \S\ref{sec:targets} \\
    \addlinespace[2pt]
    \(\Psi(x)\) & the case's aggregation of the \(K\) scores into one number & \S\ref{sec:targets} \\
    \addlinespace[2pt]
    \(R(x)\), \(p^*_c\), \(Z_c\) & the reward, the target it induces, and its normaliser & Eq.~\eqref{eq:reward}, \eqref{eq:target} \\
    \addlinespace[2pt]
    \(c\) & a condition: the control parameters the network is conditioned on & \S\ref{sec:targets} \\
    \addlinespace[2pt]
    \(P_F\), \(P_B\), \(\log Z(c)\) & forward policy, backward policy, and the learned log-normaliser head & \S\ref{sec:model} \\
    \addlinespace[2pt]
    \(\theta\), \(p_\theta(\cdot \mid c)\) & the network's learned parameters and the terminating distribution they define & \S\ref{sec:model} \\
    \addlinespace[2pt]
    \(\delta(x,c)\) & the trajectory-balance residual, Eq.~\eqref{eq:tb} & \S\ref{sec:model} \\
    \addlinespace[2pt]
    \(\lambda\) & the SubTB segment weight, 0.9 & \S\ref{sec:model} \\
    \addlinespace[2pt]
    \(L_1\) & exact distance between the policy's terminating distribution and \(p^*_c\) & \S\ref{sec:protocol} \\
    \addlinespace[2pt]
    \(n\) & sample count defining the finite-sample reference & \S\ref{sec:protocol} \\
    \addlinespace[2pt]
    \(r_s\) & Spearman rank correlation & \S\ref{sec:aggregation} \\
    \bottomrule
  \end{tabular}
\end{table}

\begin{algorithm}[htbp]
  \caption{Exact risk-composed target \(p^*_c\)}
  \label{alg:target}
  \begin{algorithmic}[1]
    \REQUIRE world \(W\) (score fields \(a_k\), nominal weights); condition \(c = (\beta_t, w_g, \text{risk block})\)
    \ENSURE exact \(p^*_c(x)\) over all \(x \in \mathcal{X}\), \(|\mathcal{X}| = H^d\)
    \FORALL{\(x \in \mathcal{X}\)}
      \STATE \(\Psi(x) \leftarrow\) the case's aggregation (Table~\ref{tab:cases})
      \STATE \(\mathrm{base}(x) \leftarrow w_g\, g(x) + w_s\, \Psi(x)\)
      \STATE \(R(x) \leftarrow \epsilon\) if \(x\) is floored (B) or vetoed (C), else \(\max(\mathrm{base}(x), \epsilon)\)
    \ENDFOR
    \STATE \(p^*_c(x) \leftarrow R(x)^{\beta_t} \big/ \sum_{x' \in \mathcal{X}} R(x')^{\beta_t}\)
  \end{algorithmic}
\end{algorithm}

\clearpage
\section{The world gate}
\label{app:gate}

Worlds are rejection-sampled, so the reported worlds are a selected sample and
the selection has to be quantified. Table~\ref{tab:gate} reports the gate's
behaviour over every world the study drew, with two independent estimates of
how often it accepts.

\begin{table}[htbp]
  \centering
  \small
  \caption{Gate behaviour over every world drawn. The last column comes from an
  ablation that skips rejection and records whether each world would have passed
  on its first draw, so it is not conditioned on passing. The two estimates
  agree.}
  \label{tab:gate}
  \begin{tabular}{@{}lccccc@{}}
    \toprule
    case & worlds & attempts mean & max & acceptance & ungated first draw \\
    \midrule
    A smooth & 152 & 1.00 & 1 & 1.000 & 1.00 \\
    B floor  & 168 & 1.00 & 1 & 1.000 & 1.00 \\
    C veto   & 136 & 1.21 & 4 & 0.829 & 0.75 \\
    D nested & 168 & 4.39 & 60 & 0.228 & 0.25 \\
    \bottomrule
  \end{tabular}
\end{table}

The gate is inert on the smooth and floor cases, mild on the veto case, and
strongly selective on the nested case, where it accepts roughly one draw in
four. Everything we conclude about Case D is therefore conditional on the gate,
which \S\ref{sec:limitations} states as a limitation.

\clearpage
\section{Protocols and definitions}
\label{app:protocols}

This appendix states the quantities \S\ref{sec:protocol} uses and the training
regimes \S\ref{sec:sparsity} compares, so that each can be reproduced from the
released code.

\paragraph{Challenge level, modes and diversity.}
The \emph{challenge level} of a world is the highest threshold on a 96-point
grid running from 0.95 down to 0 at which at least 1\% of \(\mathcal{X}\)
still satisfies the world's requirement. It is a property of that world's own
score distribution, so it is not a constant we choose per case. A \emph{mode}
is a candidate satisfying the requirement at that level, and the denominator
of mode coverage is the size of that set, reported per world. Candidates the
reward itself penalises are removed from the set before coverage is computed,
since no sampler proportional to the reward should produce them.
\emph{Diversity} is the effective number of distinct candidates,
\(\exp(H(p_S))\), where \(p_S\) is the target restricted to the satisfying set
and renormalised, and \(H\) its Shannon entropy.

\paragraph{The held-out condition grid.}
The grid is deterministic and fixed before training, with \(w_g\) held at the
middle of its range and three points per risk axis, except for Case D, which
takes two points per axis because its risk block has \(2 + O\) of them. The
resulting sizes are 27 conditions for Case A, 243 for Case B, 81 for Case C
and 192 for Case D. A condition drawn for training is rejected if its
\(L_\infty\) distance to any held-out condition falls below 0.05 in the
feature encoding the network receives.

\paragraph{The exact-KL oracle.}
The oracle shares the student's architecture and differs only in its loss. It
minimises the cross-entropy of the enumerated target,
\(-\sum_{x} p^*_c(x) \log p_\theta(x \mid c)\), averaged over the condition
batch. It therefore uses neither the \(\log Z\) head nor the sampled point
batch, and its targets are never clamped, since the clamp of
\S\ref{sec:model} is a device for the trajectory-balance residual and the
target mass it would affect is below \(e^{-25}\).

\paragraph{The five training regimes.}
All five share the network, the budget, the seed and the batch size, and
differ in the source of the off-policy half of each batch. \emph{On-policy}
has no off-policy half. \emph{Mix} draws it uniformly over \(\mathcal{X}\).
\emph{Replay} draws visited candidates with probability proportional to
\(\exp(\beta_t \log R)\) under the same clamp used in training.
\emph{Teacher} trains a second policy concurrently with trajectory balance on
a reward built from the student's trajectory-balance discrepancy over the
teacher's own samples, following \citet{kim2024}: we use their eq.~(5)
weighting, which multiplies the squared discrepancy by \(1 + C \cdot
\mathbb{1}[\delta > 0]\) with \(C = 19\), so the teacher favours candidates
the student undersamples, and we set their reward-mixing coefficient to zero.
\emph{Contrastive replay} keeps two buffers, following \citet{kim2026}: every
visited candidate goes to a positive buffer if its reward exceeds
\(\epsilon\) at the fixed condition and to a negative buffer otherwise. The
positive buffer holds unique candidates up to a capacity and evicts the
lowest-reward one; the negative buffer is first-in, first-out over recent
visits. The off-policy half is drawn from the positive buffer, and the loss
adds an InfoNCE term over balanced draws from the two buffers, scored by
\(\log P_F\).

\paragraph{The three ambiguity duals.}
With \(f\) the score vector and \(w\) the nominal weights, the KL ball uses
the entropic dual,
\[
\Phi^-(a; w, \beta, \rho) = \max_\tau \Bigl[\tau - \tfrac{1}{\beta}
  \inf_{\lambda > 0}\bigl(\lambda\rho + \lambda \log \textstyle\sum_k w_k
  e^{(\tau - a_k)^+/\lambda}\bigr)\Bigr],
\]
solved by two nested golden-section searches of 40 iterations each. The total
variation ball has a greedy closed form: a linear objective moves mass
\(\min(\rho,\, 1 - w_{\arg\max})\) onto the highest-scoring coordinate, taken
from the lowest-scoring coordinates first, with the boundary coordinate
contributing fractionally. The modified \(\chi^2\) ball
\(\sum_k (q_k - w_k)^2 / w_k \leq \rho\) has, on a candidate support \(A\)
obtained by sorting the scores ascending and taking a suffix,
\[
\sup\nolimits_A = m_A + s_A \sqrt{\rho - (1 - W_A)/W_A},
\]
with \(W_A = \sum_A w\), \(m_A\) the \(A\)-weighted mean and
\(s_A^2 = \sum_A w (f - m_A)^2\). Every suffix is checked and the feasible
maximum returned, so the result is exact. The \(\Phi^+\) variant is
\(\Phi^+(a) = -\Phi^-(-a)\) throughout.

\clearpage
\section{Sensitivity of the finite-sample reference}
\label{app:refsweep}

The reference of \S\ref{sec:protocol} is defined at a declared sample count
\(n\), and it falls as \(1/\sqrt{n}\), so a ratio to it grows as \(\sqrt{n}\).
Table~\ref{tab:refsweep} gives the student's held-out \(L_1\) as a multiple of
the reference across a range of \(n\) on the calibration family, where the
declared value is \(n = 10|\mathcal{X}| = 10{,}240\). At that value the sweep
reproduces each stored reference to five digits, 0.2227, 0.1470, 0.1981 and
0.2210.

The ordering of the four cases holds at every count we tried, smooth below
nested below veto below floor, and that ordering is what
\S\ref{sec:learnability} argues from. Whether a case clears a fixed multiple
does depend on \(n\): the floor case crosses \(3\times\) between the declared
count and twice it, and the veto case between \(20|\mathcal{X}|\) and
\(50|\mathcal{X}|\). We therefore report the endpoint together with its
declared \(n\) and give this table beside it. The sequence family is not swept
here, since one world there costs far more to enumerate repeatedly.

\begin{table}[htbp]
  \centering
  \small
  \caption{Student held-out \(L_1\) as a multiple of the finite-sample
  reference, by the reference's sample count \(n\), on the calibration family
  (\(|\mathcal{X}| = 1024\), 4 worlds per case, 20 repetitions per estimate).
  The declared count is in bold.}
  \label{tab:refsweep}
  \begin{tabular}{@{}lccccccc@{}}
    \toprule
    \(n\) & 1024 & 2048 & 5120 & \textbf{10{,}240} & 20{,}480 & 51{,}200 & 102{,}400 \\
    \midrule
    A smooth & 0.12 & 0.17 & 0.26 & \textbf{0.37} & 0.52 & 0.82 & 1.16 \\
    B floor  & 0.89 & 1.24 & 1.92 & \textbf{2.71} & 3.84 & 6.06 & 8.57 \\
    C veto   & 0.55 & 0.77 & 1.21 & \textbf{1.71} & 2.41 & 3.81 & 5.39 \\
    D nested & 0.19 & 0.27 & 0.43 & \textbf{0.60} & 0.85 & 1.34 & 1.90 \\
    \bottomrule
  \end{tabular}
\end{table}

\section{How much target mass the penalty keeps}
\label{app:epsmass}

A penalised candidate keeps reward \(\epsilon = 10^{-4}\), so it keeps target
mass, and how much depends on the temperature. Table~\ref{tab:epsmass} gives
that mass across the temperature range on both families. It is appreciable
only at the bottom of the range: 26\% of the target on the floor case at
\(\beta_t = 0.5\) on the calibration family, 7\% at \(\beta_t = 0.75\), and
negligible from \(\beta_t = 2\) upward. On the veto case it never exceeds
0.2\%, and the smooth and nested cases penalise nothing. This is why we write
\(\epsilon\)-penalised throughout: a strictly positive reward removes nothing
from the support.

\begin{table}[htbp]
  \centering
  \small
  \caption{Share of the target's probability on \(\epsilon\)-penalised
  candidates, by target temperature \(\beta_t\), averaged over 4 worlds. The
  smooth and nested cases penalise no candidate and are omitted. The
  penalised share of the space itself is given beside each case.}
  \label{tab:epsmass}
  \begin{tabular}{@{}llccccc@{}}
    \toprule
    Family & Case (space penalised) & 0.5 & 0.75 & 1.0 & 1.5 & 2.0 \\
    \midrule
    calibration & B floor (87.8\%) & 0.262 & 0.070 & 0.015 & 0.001 & 0.000 \\
                & C veto (11.4\%)  & 0.002 & 0.000 & 0.000 & 0.000 & 0.000 \\
    \addlinespace[2pt]
    deployment  & B floor (81.6\%) & 0.178 & 0.041 & 0.008 & 0.000 & 0.000 \\
                & C veto (2.6\%)   & 0.000 & 0.000 & 0.000 & 0.000 & 0.000 \\
    \bottomrule
  \end{tabular}
\end{table}

\clearpage
\section{Configuration}
\label{app:config}

Sections~\ref{sec:model} and~\ref{sec:protocol} give the settings needed to read
the results; Table~\ref{tab:config} gives the rest, so that any run can be
reconstructed. Anything not listed either took its default value or is inert in
every reported battery.


\begin{table*}[htbp]
  \centering
  \footnotesize
  \caption{Configuration. Values are read from the dataclasses that ran (\texttt{rsgfn.policy.ConditionalPolicy}, \texttt{rsgfn.train.TrainConfig}, \texttt{rsgfn.worlds.WorldConfig}, \texttt{rsgfn.conditions.ConditionRanges}), so they cannot drift from the configuration they describe. Per-battery overrides and the flags that are inert in every reported result are recorded in the released results tree.}
  \label{tab:config}
  \begin{tabular}{@{}lllp{0.46\textwidth}@{}}
    \toprule
    Group & Field & Value & Meaning \\
    \midrule
    \emph{architecture} & \texttt{cond\_dim} & 128 & condition-embedding width (FiLM conditioning input) \\
     & \texttt{dim} & 256 & trunk width \\
     & \texttt{depth} & 4 & number of FiLM residual blocks \\
     & \texttt{x1\_dim} & 64 & coordinate-embedding width \\
    \addlinespace[3pt]
    \emph{optimisation} & \texttt{loss} & tb & trajectory balance (\texttt{subtb}/\texttt{exact\_kl} in the named ablations) \\
     & \texttt{subtb\_lambda} & 0.9 & geometric segment weight, SubTB arm \\
     & \texttt{steps} & 2000 & gradient steps (overridden per battery) \\
     & \texttt{n\_conds} & 8 & conditions per batch \\
     & \texttt{n\_points} & 64 & points per condition \\
     & \texttt{uniform\_mix} & 0.5 & share of uniform-over-$\mathcal{X}$ points (off-policy mixture) \\
     & \texttt{lr} & 0.001 & Adam learning rate, trunk \\
     & \texttt{lr\_logz} & 0.01 & Adam learning rate, $\log Z$ head \\
     & \texttt{cond\_pool} & 256 & pre-priced condition pool ($\texttt{None}$ = stream fresh conditions per step) \\
     & \texttt{heldout\_radius} & 0.05 & $L_\infty$ exclusion around held-out conditions \\
     & \texttt{logit\_floor} & -25 & training-only clamp on $\beta_t \log R$; evaluation targets are never clamped \\
     & \texttt{eval\_every} & 200 & evaluation interval (steps) \\
    \addlinespace[3pt]
    \emph{world} & \texttt{H} & 32 & grid resolution per coordinate \\
     & \texttt{d} & 2 & coordinates; $|X| = H^d$ \\
     & \texttt{geometry} & grid & score-field family \\
     & \texttt{sparsity} & 1 & satisfying-set sparsity exponent \\
     & \texttt{k\_neutral} & 6 & $|S^0|$ (case A) \\
     & \texttt{k\_plus} & 5 & $|S^+|$ (cases B, C) \\
     & \texttt{k\_minus} & 4 & $|S^-|$ (case B) \\
     & \texttt{k\_named} & 3 & $|D|$ (case C) \\
     & \texttt{n\_outer} & 4 & $|O|$ (case D) \\
     & \texttt{k\_inner} & 4 & inner states per origin (case D) \\
     & \texttt{weight\_alpha} & 2 & Dirichlet concentration for all nominal weights \\
     & \texttt{gamma} & 1 & trade-off on $\Phi(S^-)$ (case B) \\
     & \texttt{floor\_quantile} & 0.35 & quantile defining the calibrated floor \\
     & \texttt{c\_named} & 0.85 & veto threshold $c_d$ (case C) \\
     & \texttt{beta\_out} & 0.5 & reference outer tail level (case D) \\
     & \texttt{rho\_out} & 0.3 & fixed outer KL radius (case D) \\
     & \texttt{hardness\_margin} & 0.05 & gate: minimum TV(risk-on, risk-off) \\
     & \texttt{veto\_frac\_range} & 0.02, 0.5 & gate: admissible veto share (case C) \\
     & \texttt{floor\_frac\_range} & 0.05, 0.7 & gate: admissible floor-fail share (case B) \\
     & \texttt{delta\_probe} & 0.1 & gate: veto-margin endpoint $\delta_{\max}$ \\
     & \texttt{max\_attempts} & 200 & gate: rejection-sampling cap \\
    \addlinespace[3pt]
    \emph{condition ranges} & \texttt{beta\_t} & 0.5, 8 & inverse-temperature range (log-uniform) \\
     & \texttt{w\_g} & 0.1, 0.9 & auxiliary-objective weight range \\
     & \texttt{rho} & 0, 1.6 & ambiguity radius range, per axis \\
     & \texttt{delta} & 0, 0.1 & veto margin range (case C) \\
    \bottomrule
  \end{tabular}
\end{table*}

\clearpage
\section{Exploration under sparsity}
\label{app:densities}

\begin{figure}[htbp]
  \centering
  \includegraphics[width=\linewidth]{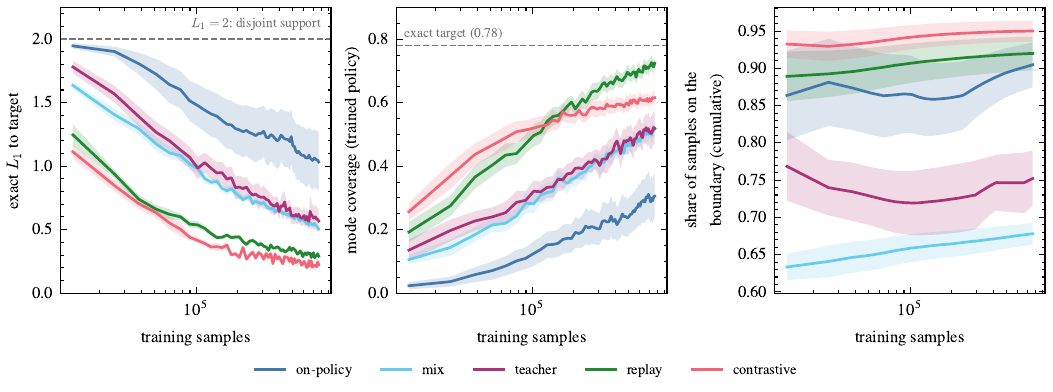}
  \caption{Training traces on the floor case at \(s{=}4\), means over 4 worlds
  \(\times\) 3 seeds with bands giving the standard error. Left: the on-policy
  regime starts at the near-disjoint ceiling \(L_1 \approx 2\) and keeps the
  widest band, which is what a trap some seeds escape and others do not looks
  like. Centre: coverage of the trained policy, against the exact target's
  0.78 at this budget. Right: the share of samples drawn one coordinate away
  from a penalised candidate.}
  \label{fig:traces}
\end{figure}

Figure~\ref{fig:traces} gives the three traces \S\ref{sec:sparsity} reads on
the floor case at \(s{=}4\): the exact \(L_1\), the mode coverage of the
trained policy, and the online diagnostic. The third is what makes the
degenerate optimum visible while training is still running. A rising share of
samples on the boundary of the penalised set, together with a coverage that
stops improving, needs no enumeration, so it is the signal a deployment can
watch once \(\mathcal{X}\) is too large to list.

Figure~\ref{fig:collapse} shows the floor case at \(s{=}4\), where the regimes
differ most. The panels below give every case at both sparsity levels, for
world 0 and seed 0, and they support three statements in \S\ref{sec:sparsity}.
At \(s{=}1\) every regime recovers the structure of the target with closely
grouped \(L_1\), so that setting does not separate them. The collapse at
\(s{=}4\) belongs to the floor case alone: on the smooth, veto and nested
cases the on-policy regime reaches an \(L_1\) comparable to the injecting
ones, because those geometries leave enough live candidates in an on-policy
batch to hold \(\log Z\) up. Across these single-world panels the ordering by
\(L_1\) is not the ordering by coverage in Table~\ref{tab:arms}, which is why
\S\ref{sec:sparsity} reads the two together and quotes means over worlds and
seeds for both.

\begin{figure}[htbp]
  \centering
  \includegraphics[width=0.92\linewidth]{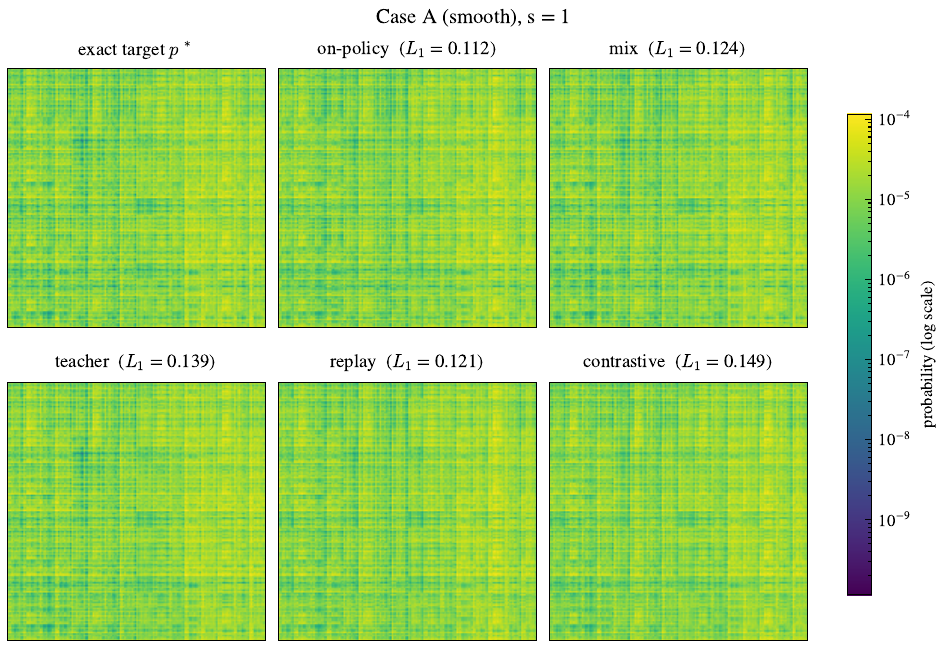}
  \caption{Case A (smooth), \(s{=}1\). Every regime recovers the target, with \(L_1\) from 0.112 to 0.149.}
  \label{fig:dens-A1}
\end{figure}

\begin{figure}[htbp]
  \centering
  \includegraphics[width=0.92\linewidth]{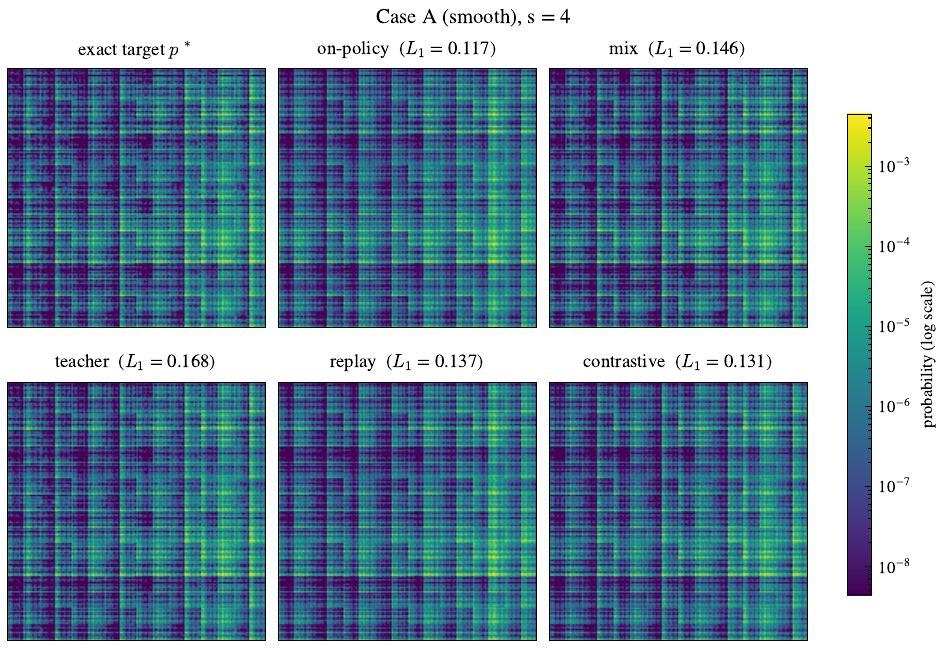}
  \caption{Case A (smooth), \(s{=}4\). Sparsity thins the target and leaves no penalised region. Every regime tracks it, and on-policy reaches the lowest \(L_1\) in this world, 0.117.}
  \label{fig:dens-A4}
\end{figure}

\begin{figure}[htbp]
  \centering
  \includegraphics[width=0.92\linewidth]{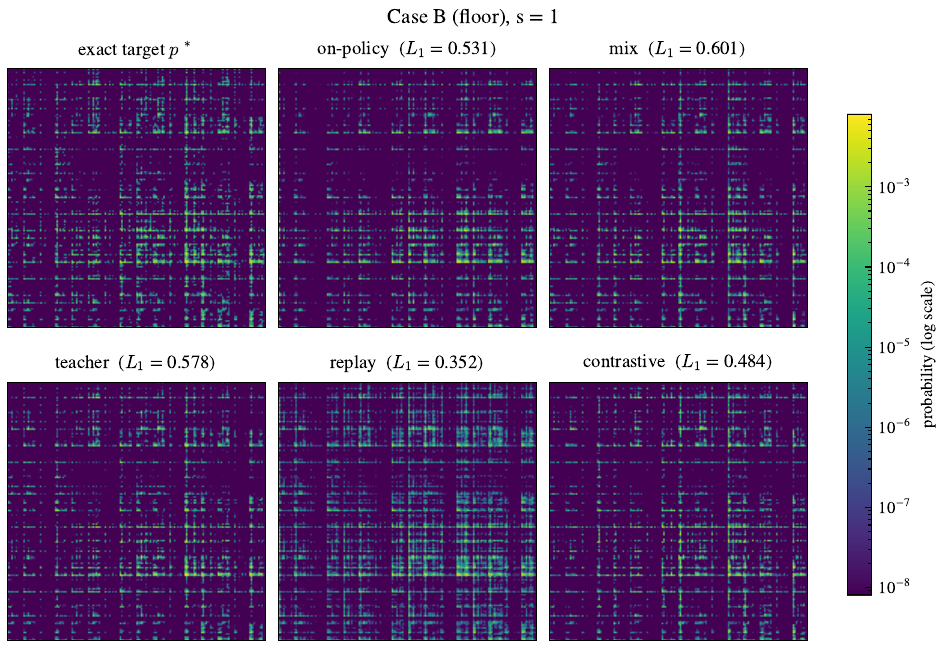}
  \caption{Case B (floor), \(s{=}1\). The floor alone already penalises 82\% of the space and the regimes separate, replay at 0.352 against on-policy's 0.531, but none of them collapses. The collapse in Figure~\ref{fig:collapse} needs sparsity on top of the floor.}
  \label{fig:dens-B1}
\end{figure}

\begin{figure}[htbp]
  \centering
  \includegraphics[width=0.92\linewidth]{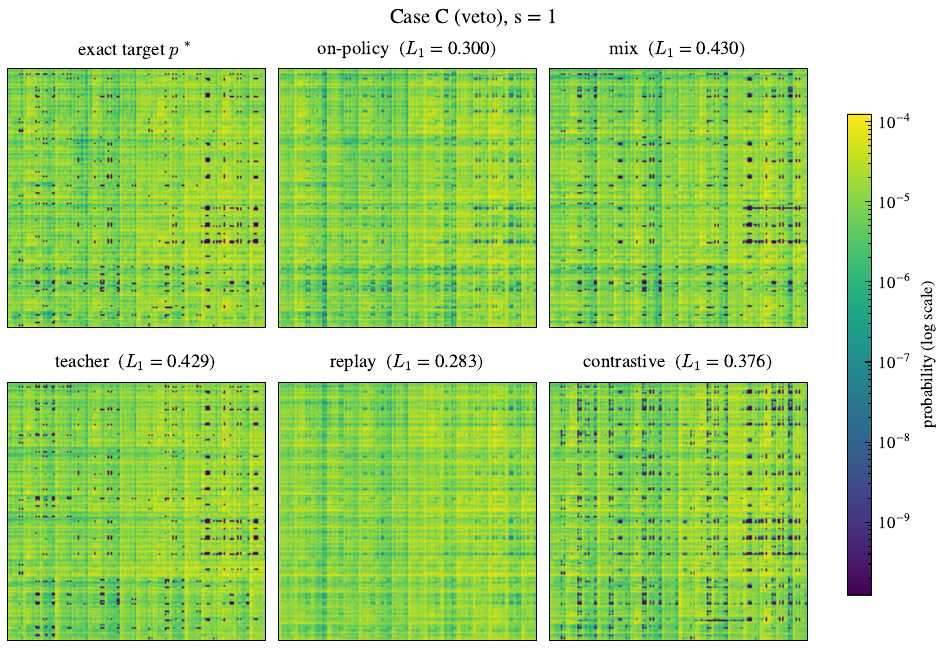}
  \caption{Case C (veto), \(s{=}1\). The scattered vetoes appear as isolated dark points in the target. Replay reaches the lowest \(L_1\) in this world, 0.283, by smoothing over exactly the structure that defines the case.}
  \label{fig:dens-C1}
\end{figure}

\begin{figure}[htbp]
  \centering
  \includegraphics[width=0.92\linewidth]{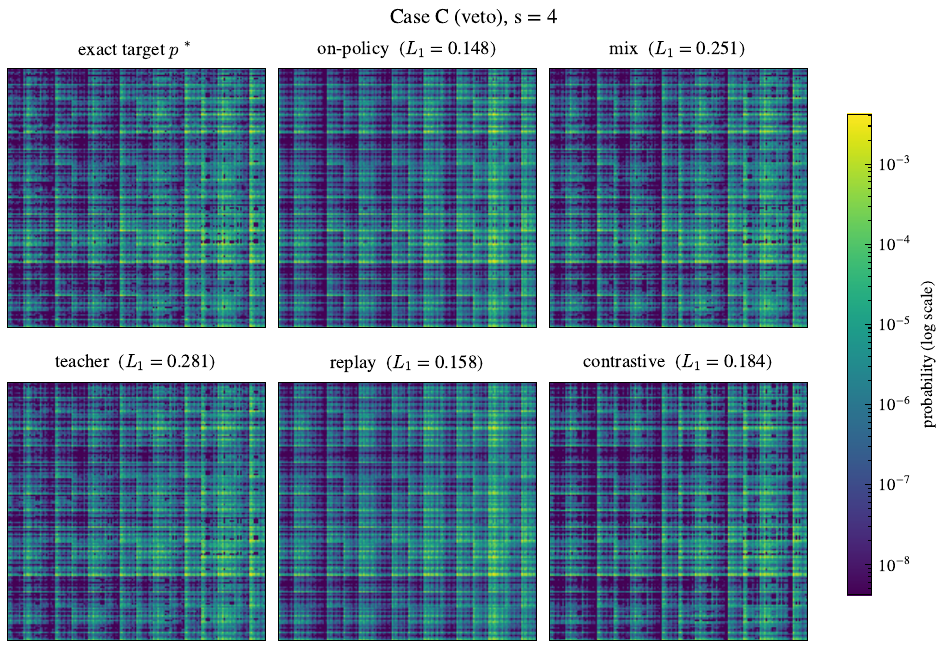}
  \caption{Case C (veto), \(s{=}4\). No collapse. On-policy reaches the lowest \(L_1\) in this world, 0.148, while its coverage averaged over worlds and seeds, 0.913, sits below three of the four off-policy regimes.}
  \label{fig:dens-C4}
\end{figure}

\begin{figure}[htbp]
  \centering
  \includegraphics[width=0.92\linewidth]{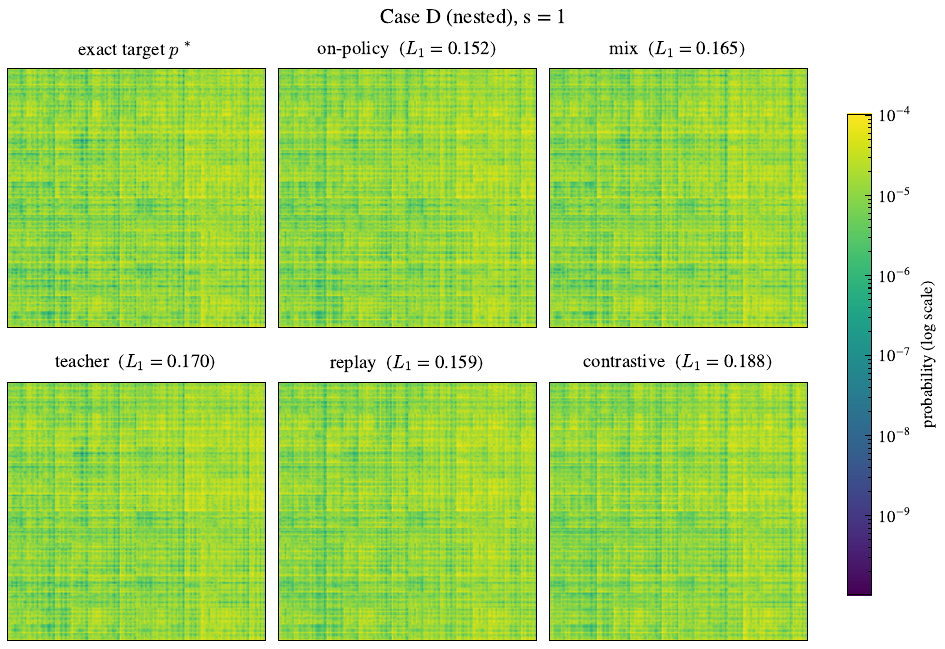}
  \caption{Case D (nested), \(s{=}1\). The nested target is dense and every regime matches it closely, with \(L_1\) from 0.152 to 0.188.}
  \label{fig:dens-D1}
\end{figure}

\begin{figure}[htbp]
  \centering
  \includegraphics[width=0.92\linewidth]{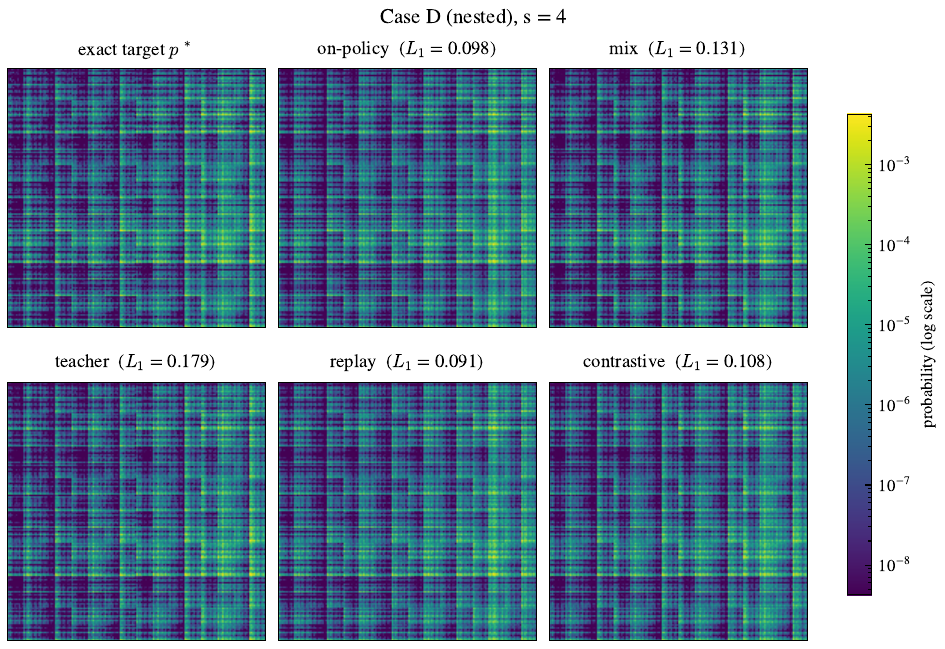}
  \caption{Case D (nested), \(s{=}4\). Sparsity introduces banded penalised regions and no collapse; replay reaches the lowest \(L_1\) in this world, 0.091. Its coverage in Table~\ref{tab:arms} is 0.518, below both injecting regimes, so on this case the two measurements order the regimes differently.}
  \label{fig:dens-D4}
\end{figure}

\section{Difficulty axes and their cost}
\label{app:prices}

Section~\ref{sec:price} names four difficulty axes and the case that binds
each one, and Table~\ref{tab:prices} collects them. The entries separate the
axes by character: the cost along the size axis does not flatten with resolution, because the
veto boundary grows with the space and has to be visited. The cost along the
depth axis is measured while the student's error is still falling, which is
what makes a longer budget recover it.

\begin{table}[htbp]
  \centering
  \small
  \caption{Held-out \(L_1\) as a multiple of the finite-sample reference, per difficulty axis, with the case that binds each one.}
  \label{tab:prices}
  \begin{tabular}{@{}p{3.0cm}p{5.6cm}p{5.6cm}@{}}
    \toprule
    Axis & cost (binding case) & what recovers it \\
    \midrule
    size, \(H\!: 16 \to 64\) & veto boundary grows: C 1.4\(\times\) \(\to\) 2.1\(\times\) & none within budget: more resolution is more boundary \\
    \addlinespace[2pt]
    depth, \(d\!: 2 \to 4\) & B breaches: 2.7\(\times\) \(\to\) 3.6\(\times\) & tripling the budget returns it below 3\(\times\) \\
    \addlinespace[2pt]
    peaked weights & B 3.2\(\times\), unmoved by budget & open failure \\
    \addlinespace[2pt]
    score noise \(\sigma\) & B 5.0\(\times\), C 3.2\(\times\), and B's \(\sigma\)-oracle itself above 3\(\times\) & nothing we tried \\
    \bottomrule
  \end{tabular}
\end{table}

\end{document}